# Precursors, Pathways, and State-Dependent Reliability of Long-range ENSO Prediction in the CESM2 Seasonal-to-Multiyear Large Ensemble

Yudi Mao[1], Gan Zhang[1*], Hui Li[2], Dillon J. Amaya[3],

Stephen G. Yeager[2], and Sen Zhao[4]

[1] *Department of Climate, Meteorology & Atmospheric Sciences, University of Illinois Urbana-Champaign, Urbana, Illinois*

[2] *Climate and Global Dynamics Laboratory, NSF National Center for Atmospheric Research (NCAR), Boulder, Colorado*

[3] *Department of Marine, Earth and Atmospheric Sciences, North Carolina State University, Raleigh, North Carolina*

[4] *Department of Atmospheric Sciences and International Pacific Research Center, School of Ocean and Earth Science and Technology, University of Hawai'i at Mānoa, Honolulu, Hawai'i*

**Corresponding author*: Gan Zhang, gzhang13@illinois.edu

ABSTRACT

Understanding the sources of ENSO prediction skill and error is essential for improving long-range climate prediction. Here we examine ENSO precursors in CESM2 Seasonal-to-Multiyear Large Ensemble (SMYLE) hindcasts initialized in February and August during 1970–2019, using ensemble sensitivity analysis, skill-stratified composites, and an extended nonlinear recharge oscillator (XRO) benchmark. SMYLE exhibits meaningful long-lead skill, with February-initialized predictions remaining skillful through the following winter, and August-initialized predictions retaining useful skill at the December target sixteen months ahead. However, ensemble mean skill metrics conceal off-setting state-dependent errors. Relative to a resampled calibration null, the ensemble is overconfident for strong ENSO events while exhibiting a warm bias for weak events. February predictability arises from canonical Bjerknes feedback dynamics reinforced by subtropical–tropical interactions involving the North Pacific Meridional Mode (NPMM). The best February-initialized predictions begin from a coherent recharged La Niña state including warm western Pacific subsurface anomalies, with equatorially confined anomalies and relatively weak sensitivities to extratropical influences thereafter. Second-year ENSO predictability in August-initialized predictions instead emerges through a delayed pathway characterized by inter-basin preconditioning. Skillful cases begin with a cold tropical equatorial South Atlantic and a warmer tropical western Indian Ocean, while Pacific processes play a reduced role until the following year. SMYLE matches the skill of the cross-validated XRO at both December targets. While SMYLE's sensitivities agree with the observationally fitted benchmark, the Atlantic couplings appear misrepresented. These findings suggest that improving the initial Pacific state and inter-basin coupling fidelity in GCMs may help improve ENSO prediction.

## 1. Introduction

The El Niño–Southern Oscillation (ENSO) is the dominant mode of interannual climate variability, influencing global weather extremes, ecosystems, and socioeconomic systems (Philander 1983; McPhaden et al. 2006). Improving ENSO prediction is therefore central to seasonal-to-interannual climate services. Yet despite decades of development, the skill of dynamical ENSO predictions with global climate models (GCMs) appears to have plateaued or even declined recently (Barnston et al. 2012; Xue et al. 2013; Pegion et al. 2020). The challenge appears partly associated with the spring predictability barrier (SPB), a rapid decline in forecast skill when predictions cross boreal spring (Torrence and Webster 1998; Webster and Yang 1992). The skill plateau may arise if operational prediction systems operate near the intrinsic predictability limit of tropical ocean (Newman and Sardeshmukh 2017). Nonetheless, growing evidence suggests that aspects of ENSO remain predictable beyond the 6–12-month horizon (Wu et al. 2021), and machine learning and reduced order models have reported useful skill at leads approaching 18 months (Ham et al. 2019; Zhao et al. 2024; Wang et al. 2026). Together, these results motivate a focused diagnosis of where dynamical prediction systems gain and lose ENSO skill, especially for long-lead predictions.

ENSO dynamics are primarily governed by coupled ocean–atmosphere feedbacks in the tropical Pacific, particularly the Bjerknes feedback linking sea surface temperature (SST), trade winds, and thermocline variations (Bjerknes 1969; Jin 1997). Meanwhile, equatorial heat content provides the subsurface memory that underpins predictability through recharge–discharge adjustment (Jin 1997). A growing body of research further suggest that ENSO evolution and predictability are shaped by interactions with remote climate variability, including extratropical Pacific modes and variability in other basins (e.g., Kayano et al. 2011; Amaya 2019; Jeffree et al. 2026). Within the Pacific, the North Pacific Meridional Mode (NPMM) can trigger and amplify ENSO through subtropical–tropical teleconnections that modulate trade winds and the thermocline (Chang et al. 2007; Amaya 2019; Li et al. 2025). Recent hindcast experiments indicate that the NPMM-associated atmospheric variability exerts a more sustained influence on ENSO development than intraseasonal westerly wind forcing (Liang et al. 2025). Beyond the Pacific, Atlantic SST anomalies associated with the Atlantic Equatorial Mode (AEM; a.k.a. Atlantic Niño) and tropical North Atlantic variability can influence Pacific convection and surface winds, potentially modulating ENSO onset (e.g., Saravanan and Chang 2000; Kayano et al. 2011; Ham et al. 2013). In the Indian Ocean, both the basin-wide mode and the Indian Ocean Dipole can modulate ENSO development via inter-basin interactions (e.g., Wu and Kirtman 2004; Izumo et al. 2010; L. Zhang et al. 2021; Jin et al. 2023). Despite the substantial progress, it remains challenging to establish whether GCM-based forecast systems represent these interactions correctly and whether their inclusion truly contributes to enhanced long-lead skill.

The Seasonal-to-Multiyear Large Ensemble (SMYLE) prediction system (Yeager et al. 2022) provides an opportunity to address these questions. While earlier multi-model ensembles, such as the North American Multi-Model Ensemble (NMME; Kirtman et al. 2014) and the ENSEMBLES project (Weisheimer et al. 2009), have demonstrated improved climate prediction skill, these systems combine models with differing physics and initialization strategies. This makes attributing forecast errors to specific sources challenging. In comparison, a single-model system avoids this attribution ambiguity by holding physics and initialization fixed across members. The SMYLE prediction system, which is based on a coupled Earth system model (CESM2), makes a combination of design choices including 24-month forecast range, 20-member ensembles, and 50-year hindcast record (Yeager et al. 2022). These design choices, together with a rich archive of atmosphere–ocean variables, makes SMYLE particularly well suited for isolating the precursors and pathways of ENSO prediction skill. The 24-month forecast range, for example, is shared by few prediction systems and enables the exploration of long-range predictability. The 20-member ensemble affords probabilistic evaluation — including reliability and its dependence on the ENSO state — and process-oriented analyses of the modeled ENSO evolution.

Direct initial-condition sensitivity experiments are prohibitively expensive in coupled GCMs, but the success of linear frameworks in ENSO prediction (Newman and Sardeshmukh 2017; Wang et al. 2026) suggests linear diagnostics capture much of the relevant sensitivity. We therefore adopt ensemble sensitivity analysis (ESA) (Ancell and Hakim 2007; Torn and Hakim 2008), a statistical technique that quantifies relationships between forecast-state anomalies and forecast outcomes across ensemble members. ESA has been applied successfully to analyze high-impact weather events (e.g., Ancell and Coleman 2022). A pilot study using NMME hindcasts suggested its promise for diagnosing ENSO forecast errors (Zhang 2023), though that study was limited due to data availability challenges. Here we apply ESA to the forecasts of SMYLE (Yeager et al. 2022) and stratify the hindcasts by levels of forecast success. We also benchmark the SMYLE against the extended nonlinear recharge oscillator model (XRO) (Zhao et al. 2024) under both full-fit and cross-validated evaluation. The XRO has demonstrated skill beyond the spring predictability barrier and provides an observationally constrained reference whose individual SST modes can be ablated, enabling process-level attribution that is challenging with a coupled GCM. We focus on predictions initialized in February and August to trace sources of predictability and identify errors in first-year and second-year ENSO predictions, respectively. Specifically, we address three research questions:

1. How do the skill and reliability of long-range ENSO predictions from CESM2-SMYLE depend on ensemble size and ENSO state?

2. What precursors and dynamical pathways affect ENSO prediction skill in February- and August-initialized SMYLE hindcasts?

3. Are the coupled ocean–atmosphere processes in SMYLE consistent with the observationally constrained XRO?

The remainder of the paper is organized as follows. Section 2 describes the SMYLE hindcasts, observational datasets, and analysis methods, including the ESA framework. Section 3 provides an overview of ENSO forecast skill, its dependence on ensemble size, and the state dependence of ensemble reliability. Section 4 examines February-initialized predictions and the mechanisms shaping first-year ENSO evolution and forecast success. Section 5 analyzes August-initialized predictions focusing on how the second-year ENSO predictability emerges through delayed dynamical pathways and inter-basin preconditioning. Section 6 compares the skill and coupled processes of SMYLE with those of the XRO. Section 7 summarizes the main findings and discusses limitations.

## 2. Model experiments and analysis methods

### *a. SMYLE Hindcasts*

This study uses initialized hindcast simulations from the SMYLE prediction system, consisting of 24-month ensemble forecasts conducted with the Community Earth System Model version 2 (CESM2) at a nominal 1° horizontal resolution in all model components (Yeager et al. 2022). The CESM2 configuration follows Danabasoglu et al. (2020), with the atmosphere represented by the Community Atmosphere Model version 6 (CAM6), the ocean by the Parallel Ocean Program version 2 (POP2) with 60 vertical levels, sea ice by CICE version 5.1.2, and land processes by the Community Land Model version 5 (CLM5) with interactive biogeochemistry. The SMYLE hindcasts are initialized on the first day of November, February, May, and August for each year from 1970 to 2019. Each initialization consists of 20 ensemble members, with ensemble spread generated through random perturbations applied to the initial atmospheric state (Richter et al. 2020, 2022). This study focuses on February- and August-initialized forecasts to contrast ENSO predictability across different seasonal initialization regimes. While our analyses of February-initialized predictions emphasize first-year ENSO evolution, the analyses of August-initialized predictions mainly target the sources of predictability at second-year lead times. Forecast lead is counted in months, with Lead 0 denoting the monthly mean of the initialization month. As an example, the target December Niño-3.4 index corresponds to Lead 10 for February forecasts.

We mainly analyze monthly mean model output fields, including SST, subsurface ocean temperature, sea level pressure (SLP), and surface winds. Except for the calculation of the Niño-3.4 index, all the model fields of SMYLE are remapped from the native ocean grid to a 2° × 2° grid to reduce computational cost while preserving the large-scale variability relevant to ENSO diagnostics. Remapping across the seam of the dipolar POP2 grid introduces a narrow band of spurious values near 40°W (317°–322°E) in the regridded fields. We mask this band in all remapped ocean fields and refill it by zonal interpolation. The masked band lies outside all regional index domains used by this study (Table 1). To account for lead-time-dependent model drift, monthly anomalies are computed relative to a hindcast climatology over 1981–2010, calculated separately for each initialization month and forecast lead time.

*b. XRO Model*

The extended nonlinear recharge oscillator model (XRO; Zhao et al. 2024) is fitted to the Ocean Reanalysis System 5 (ORAS5) over 1970–2021 (Zuo et al. 2019). The dataset choice is mainly motivated by the need for ocean heat content information and the extended year range. Following Zhao et al. (2024), we first work with ten oceanic variables, which includes two ENSO state indices and eight remote SST indices. The indices here are identical to Zhao et al. (2024) except for the equatorial Atlantic index. Instead of using the equatorially confined Atlantic Niño index, we use the SST in the tropical equatorial south Atlantic (TESA; 35°W – 0°, 15°S – 5°N) to accommodate the coarser resolution of our SMYLE analysis. Sensitivity tests suggest the impact of this change on the XRO model skill is small.

We fit several versions of XRO models. The standard version generates 20-member stochastic ensembles, matching the SMYLE ensemble size. A comparison of the standard version with the 200-member version helps estimate the sampling penalty incurred by the ensemble size. A leave-two-years-out cross-validated (LTYCV) version tests how much of the standard version's skill reflects training on data that overlap the forecast window. To fit the LTYCV version that makes prediction of a two-year period, we remove the target period from the training record and fit the operator and noise statistics on the remainder years. Two years are withheld to keep the December target in our analyses out of training, and whole calendar years are removed so that every retained month keeps its position in the seasonal cycle. The observed initial state is retained throughout experiments, so the LTYCV assesses the fitted model and not the initialization.

We conduct ablation tests to quantify the impacts of inter-basin coupling on predictions. For each test we set the coupling from one predictor into the ENSO tendency equations to zero in all twelve calendar months and repeat the reforecasts. Every other element of the fitted model is unchanged, including the ENSO-to-predictor feedback and the couplings among predictors, so the test isolates the direct forcing pathway. Since indirect pathways through other predictors remain active, the resulting skill change is therefore a lower bound on a predictor's total influence. Because only the deterministic operator is altered, the residual variance and lag-1 autocorrelation that generate the stochastic forcing can be kept unchanged. At each initialization, we draw the identical forcing for the full and ablated forecasts. Skill differences therefore reflect the removed coupling rather than a different realization of stochastic forcing. Skill is scored on the 20-member ensemble mean, and $\Delta ACC = ACC_{full} - ACC_{ablated}$ is evaluated on linearly detrended Niño-3.4 anomalies across the 50 initializations of 1970–2019. Confidence intervals come from a bootstrap (N=5000) that resamples verification years and scores both configurations on the same resampled years. We also fit a version retaining tropical Pacific interactions only, obtained by zeroing the coupling of all eight remote predictors into the ENSO equations. This leaves the tropical Pacific subsystem closed but with coefficients and noise statistics inherited from the full ten-variable fit. Finally, we construct a sensitivity test by adding an SST index of the North Atlantic Subpolar Gyre (NASPG). The method and results are further discussed in Supplementary Materials.

*c. Observational Data and Forecast Verification*

SMYLE and XRO forecasts are verified against the Hadley Centre Sea Ice and Sea Surface Temperature dataset (HadISST) (Rayner et al. 2003), which provides monthly SST on a 1° × 1° grid from January 1870 to May 2024. ENSO variability is quantified using the Niño-3.4 index, the area-mean SST anomaly over 170°W–120°W and 5°S–5°N. We also analyze the regional domains that represent well-documented ENSO precursor regions, including subtropical Pacific modes and cross-basin influences. Besides the regional indices used by Zhao et al. (2024), we also examine other regions and variables (e.g., SLP) of interest, as summarized by Table 1.

We evaluate ENSO forecast skill using the anomaly correlation coefficient (ACC), root-mean-square error normalized by the observed interannual standard deviation (nRMSE), and the fair (ensemble-size-unbiased) Continuous Ranked Probability Score (CRPS), each computed as a function of forecast lead month. At each lead, anomalies of forecast and observed values valid in the same calendar month are calculated by removing their respective 1970–2019 means and linearly detrended, so that skill reflects interannual variability rather than long-term trends. To assess the probabilistic reliability of forecasts, we also introduce the ratio of forecast spread to nRMSE, as well as rank histograms. The sensitivity of skill to ensemble size is quantified by a member bootstrap: for ensemble sizes of 1, 3, 10, and 20, members are resampled with replacement 200 times, and the 5th–95th percentile range across draws provides the uncertainty estimate.

*d. Ensemble Sensitivity Analysis*

The ESA is modified to diagnose the sensitivity of forecasts to previous states (Ancell and Hakim 2007; Hakim and Torn 2008). For an ensemble of size $M$, the sensitivity of the ensemble mean value of a forecast metric $J$ (e.g., the December Niño-3.4 index anomaly) to an initial state variable $x$ (e.g., SST anomaly at a given grid point) is given by:

$$\frac{\partial J}{\partial x} = \frac{cov(J,\ x)}{var(x)} \qquad (1)$$

Here, $x$ and $J$ are $1 \times M$ vectors of ensemble perturbations of the state variable and forecast metric, respectively, with the ensemble mean removed. Meanwhile, *cov* denotes their covariance, and *var* denotes the ensemble variance of $x$. Equation (1) is a linear regression across ensemble members in which the independent variable is the state anomaly at a given grid point, and the dependent variable is the forecast metric.

In the classical formulation $x$ is the analysis at initialization, so that (1) measures the sensitivity of the forecast outcome to initial-condition uncertainty (Ancell and Hakim 2007; Hakim and Torn 2008). Here we generalize $x$ to the forecast state at any lead $\ell$ preceding the target lead L, so that $x = x(\ell)$ for $\ell = 0, \ldots, L - 1$, with $J$ the target December Niño-3.4 anomaly (Lead 10 for February-initialized forecasts; Lead 16 for August-initialized forecasts). The classical case is recovered at $\ell = 0$. For $\ell > 0$, $x$ and $J$ are both forecast states, and (1) quantifies the degree to which anomalies arising within the ensemble at lead $\ell$ covary with the eventual target anomaly, tracing how the predictable signal organizes along the forecast trajectory.

We report a correlation-based form of (1), which facilitates comparison across variables with different units.

$$r(\ell) = \frac{cov(J,\ x(\ell))}{var(J) \cdot var(x(\ell))} \quad (2)$$

so that the correlation is the regression sensitivity normalized by the ratio of ensemble standard deviations. ESA maps show these per-year correlation patterns averaged over all initialization years (1970–2019), unless the years are otherwise specified. Grid point significance tests whether the sensitivity is sign-consistent across initialization years. Per-year correlations are Fisher-z transformed and evaluated with a one-sample t-test across years, and stippling indicates significance after Benjamini–Hochberg false-discovery-rate control ($\alpha = 0.05$). Wind vectors are shown only where at least one component is significant at $p < 0.05$.

A caveat governs the interpretation of ESA at short leads. Classical ESA works with the observationally constrained estimate of the initial state across ensemble members, so the resulting sensitivities indicate how errors in that estimate would propagate into the forecast (Ancell and Hakim 2007). However, SMYLE's ensemble spread is generated by atmospheric initial perturbations alone, and all members are initialized with the same ocean analysis (Yeager et al. 2022). At the shortest leads, the within-ensemble variance of oceanic fields therefore reflects a few weeks of internally generated divergence rather than initial-condition uncertainty. Consequently, our sensitivities describe how coupled anomalies generated internally within the ensemble covary with the forecast outcome, not how uncertainty in the initial ocean state maps onto forecast error. Initial state contrasts between well- and poorly-predicted years are instead established from ensemble mean composites (Section 2e).

*e. Composite Analysis Based on Prediction Skill*

We examine the initial state contrasts between the most and least skillful ENSO predictions using ensemble mean composites. Specifically, we build composites by ranking all years by the absolute error between the ensemble mean and observed December Niño-3.4 anomaly (Lead 10 for February; Lead 16 for August). The top and bottom eight years define the "best-predicted" and "worst-predicted" cases, used consistently throughout the composite, ESA, and regional-index analyses (Table 1). Composite differences (best minus worst) are formed from ensemble mean anomaly fields, and their significance is assessed with a two-tailed Welch's t-test at each grid point and lead. For completeness, we denote signals at $p < 0.10$ and $p < 0.05$ in these composite analyses.

Composite and index anomalies are computed relative to the 1981–2010 hindcast climatology without detrending, whereas skill metrics are detrended (Section 2c). We verified that the linear trend projects only weakly onto the best-minus-worst composites and that all composite index contrasts retain their sign and significance when recomputed on detrended anomalies. With only eight members in each composite group, false discovery rate control removes nearly all grid point signals regardless of their spatial coherence. Grid point significance in the composite difference maps is thus shown at raw two-sided Welch's-t thresholds and is intended as descriptive guidance to spatial structure. Confirmatory inference instead rests on the small set of pre-defined regional indices (Table 1). We verify their best–worst contrasts to be robust to the ranking metric (e.g., CRPS), to composite size (N = 6–12), and to detrending.

*Table 1. Top eight best- and worst-predicted years by initialization month (1970–2019), and regional climate indices used in SMYLE composite analyses. The superscripts after individual years indicate the initial state (first symbol) and the target state (second symbol) of the ENSO, with + for El Niño, − for La Niña, O for neutral. Index names with stars (★) indicate that the indices are also used as an XRO predictor mode (Zhao et al. 2024). Dipole indices are defined as western-pole minus eastern-pole area mean. North Pacific Meridional Mode (NPMM), South Pacific Meridional Mode (SPMM), Indian Ocean Basin (IOB), Indian Ocean Dipole (IOD), Subtropical Indian Ocean Dipole (SIOD), Tropical North Atlantic (TNA), Tropical Equatorial South Atlantic (TESA), South Atlantic Subtropical Dipole (SASD), North Atlantic Subpolar Gyre (NASPG), West Pacific Subsurface temperature anomalies (Wsub, averaged over the 0–300 m layer), East Pacific SLP (Eslp), and Northeast Pacific SLP (NEslp).*

| | **February Initialization** | **August Initialization** |
|---|---|---|
| **Best-Predicted Years (8)** | $1982^{O+}$, $1989^{-O}$, $2001^{-O}$, $2012^{-O}$, $2000^{--}$, $1988^{O-}$, $1985^{-O}$, $2008^{--}$ | $1983^{O-}$, $1992^{OO}$, $1977^{OO}$, $1979^{OO}$, $1980^{OO}$, $2012^{+O}$, $1975^{-+}$, $2011^{-O}$ |
| **Worst-Predicted Years (8)** | $1994^{O+}$, $1975^{O-}$, $2017^{O-}$, $1991^{O+}$, $1980^{OO}$, $2009^{-+}$, $2003^{+O}$, $1979^{O+}$ | $2009^{+-}$, $1993^{O+}$, $1974^{O-}$, $1996^{O+}$, $2006^{O-}$, $1971^{-+}$, $2014^{O+}$, $2008^{O+}$ |
| **Index** | **Spatial Domain** | |
| **NPMM ★** | 160°W – 120°W, 10°N – 25°N | |

| SPMM ★ | 110°W – 90°W, 25°S – 15°S |
|---|---|
| IOB ★ | 40°E – 100°E, 20°S – 20°N |
| IOD *(W−E)* ★ | W: 50°E–70°E, 10°S–10°N, E: 90°E–110°E, 10°S–0° |
| SIOD *(W−E)* ★ | W: 65°E–85°E, 25°S–10°S, E: 90°E–120°E, 30°S–10°S |
| TNA ★ | 55°W – 15°W, 5°N – 25°N |
| TESA ★ | 35°W – 0°, 15°S – 5°N |
| SASD *(W−E)* ★ | W: 60°W–0°, 45°S–35°S, E: 40°W–20°E, 30°S–20°S |
| NASPG | 60°W – 20°W, 40°N – 60°N |
| Wsub Pacific | 120°E – 170°E, 10°S – 20°N |
| Eslp Pacific | 135°W – 80°W, 10°S – 10°N |
| NEslp Pacific | 140°W – 120°W, 30°N – 45°N |

## 3. ENSO Prediction Skill of CESM2 SMYLE

While the ENSO prediction skill of SMYLE has been evaluated previously (Yeager et al. 2022), we briefly discuss key deterministic skill metrics before focusing on the probabilistic prediction skill of SMYLE. Following common practice in ENSO forecast verification, the discussion below emphasizes the ACC=0.5 reference level. The long-lead skill statements do not hinge on that convention. Besides the basic skill analyses, the discussion here quantifies the sensitivity of prediction skill to ensemble size and provides context for the other diagnostic analyses. Figure 1 shows the ACC, nRMSE, fair CRPS, and spread–nRMSE ratio for Niño-3.4 predictions over 1970–2019.

SMYLE exhibits meaningful long-lead skills in predicting ENSO events (Fig. 1). For the 20-member ensemble mean, February-initialized forecasts sustain an ACC above 0.5 continuously through lead month 14 (March of Year-2), capturing the full evolution of the first-winter ENSO event. August-initialized forecasts start from lower skill and decline through the boreal spring of Year-2 (Figs. 1b,d), but their skill re-emerges in the second forecast winter. After dipping to 0.43–0.48 at Leads 11–13, the ACC recovers to ≈ 0.52 at Leads 14–16, so that useful skill extends through December of Year-2, the second-winter ENSO peak itself. This prediction horizon exceeds the typical ~12-month limit of many operational ENSO forecast systems (Pegion et al. 2020). While SMYLE skill is somewhat lower than that achieved by recent machine-learning-based and physics-based conceptual ENSO prediction models (Ham et al. 2019; Zhao et al. 2024), these results demonstrate that dynamical GCMs can attain skillful ENSO predictions at multi-seasonal to interannual lead times.

Ensemble size strongly affects the ACC and nRSME. Increasing the ensemble size from 1 to 20 members systematically decreases nRMSE and increases ACC, with particularly strong benefits for August-initialized forecasts. For February initializations, the ACC during lead months 3–5 (April–June) increases from 0.74 (single member) to 0.84 (20 members) (Fig. 1a). The benefit is larger for August initializations, with the mean ACC rising from 0.42 to 0.51 over Leads 11–16 (i.e., June–November of Year-2; Fig. 1b). Similar skill improvements related to the ensemble size increase are evident in nRMSE (Figs. 1c,d), especially near the SPB periods. These improvements reflect the suppression of atmospheric internal variability by ensemble averaging, consistent with the interpretation of the SPB as a period of enhanced atmospheric noise and weakened ocean–atmosphere coupling during boreal spring (Webster and Yang 1992; Torrence and Webster 1998; Hu and Duan 2016; Levine and McPhaden 2015; Mukhin et al. 2021). Marginal gains become small beyond about 10 members for both initializations, indicating that a 20-member ensemble captures most of the ensemble mean skill.

The probabilistic metrics reveal lead-dependent calibration variations. The spread–nRMSE ratio is well below unity in the first months after initialization (Figs. 1g–h), suggesting the ensemble is markedly under-dispersive while it remains tightly constrained by the initial state. This is an expected consequence of the atmosphere-only perturbation strategy, since oceanic spread must develop through coupled dynamics (Section 2d). In later forecasts, the ratio approaches the ideal value of one as forecasts pass through the SPB, when stochastic atmospheric variability inflates the spread toward the actual forecast uncertainty. Consistently, the fair CRPS does not grow monotonically with lead but plateaus or dips near the boreal spring windows (Figs. 1e–f), indicating that the probabilistic penalty of growing ensemble mean error is partly offset by improved ensemble spread.

Aggregate calibration statistics, however, conceal a strongly state-dependent miscalibration (Fig. 2). Conditional rank histograms stratified by the observed Niño-3.4 anomaly magnitude show a pronounced U-shape for strong events ($|\Delta T| \geq 1$ °C). Aggregated across all leads, the outermost four rank bins receive 46.5% (February, N = 275) and 41.9% (August, N = 272) of counts, while the three central bins hold only 8.0% and 9.9% (Fig. 2). Weak events, by contrast, show relatively even spreads but exhibit a systematic warm bias, with observations falling in the lower half of the ensemble far more often than expected. Because stratifying by the observed anomaly conditions on the outcome, a U-shaped conditional histogram is expected even for a perfectly calibrated ensemble. We therefore compare against a resampled calibration null in which a randomly chosen member serves as the verifying observation under the same selection. The observed outer-bin frequencies (46.5% and 41.9%) significantly exceed the null expectation of ≈32% ($p = 0.004$ and $0.026$), and the weak-stratum warm bias likewise exceeds the null ($p \leq 0.002$). Since the overconfident forecasts of strong ENSO events and biased forecasts of weak events offset in aggregate, the overall spread–nRMSE ratio (Figs. 1g–h) and rank histograms without the ENSO stratification (Fig. S1) mask this state dependence.

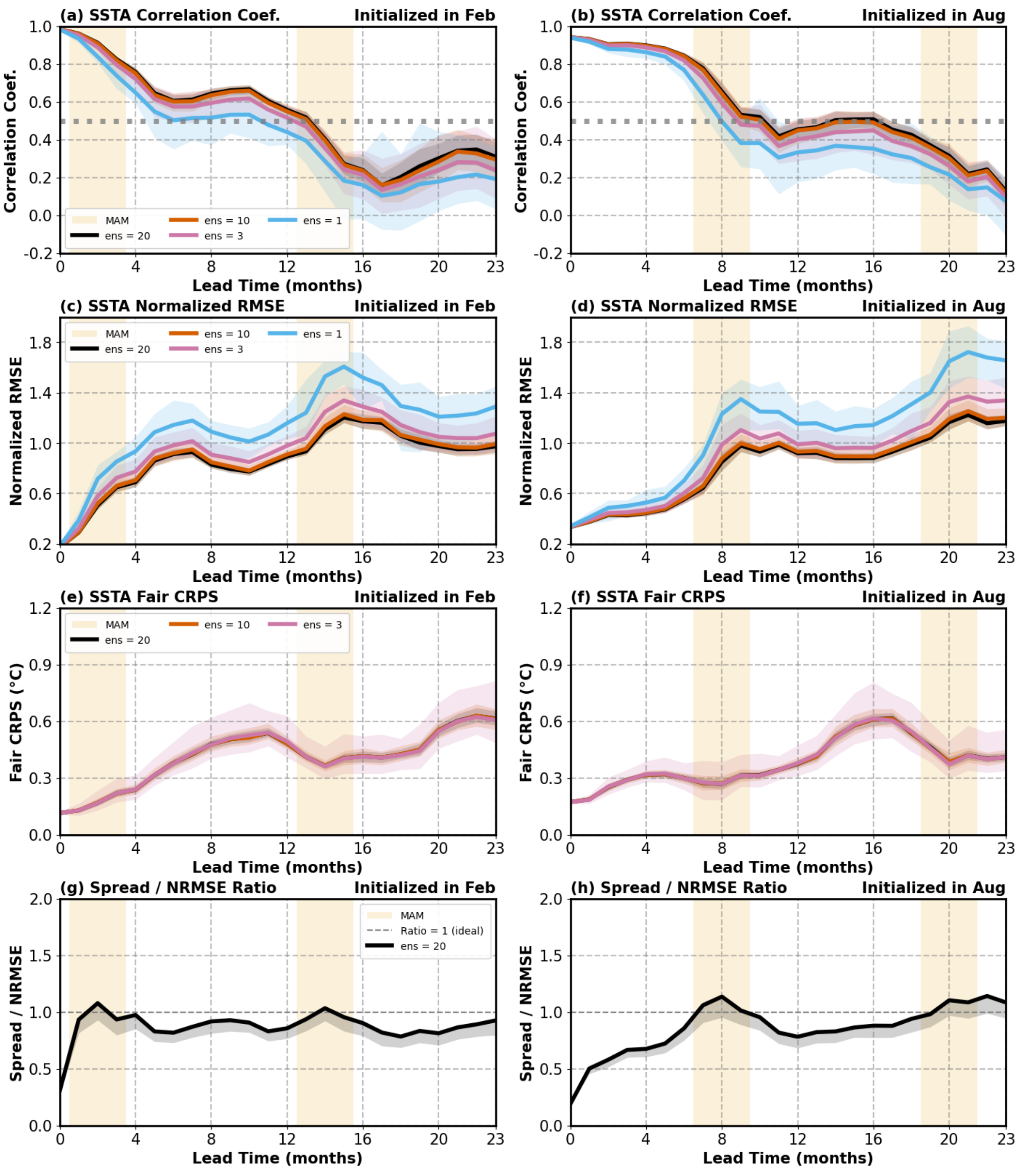

*Fig. 1. The SMYLE Niño-3.4 SST forecast skill as a function of lead time for February (left) and August (right) initializations over the 1970–2019 hindcast period. (a)–(b) Anomaly correlation coefficient (ACC); (c)–(d) normalized root-mean-square error (nRMSE), defined as RMSE divided by the lead-specific standard deviation of the linearly detrended observed anomaly; (e)–(f) fair continuous ranked probability score (CRPS, °C); (g)–(h) normalized spread–nRMSE ratio for the full 20-member ensemble. In (a)–(f), results are shown for four ensemble sizes (1, 3, 10, and 20 members). Shading indicates the 90% confidence interval derived from 200 bootstrap draws, each randomly selecting ensN members with replacement. The dottedl lines in (a) )–(b) indicate ACC=0.5, a common threshold for skillful predictions. In (e)–(f ), fair CRPS is not defined for a single member and is therefore omitted for ens = 1. In (g)–(h), the spread is the mean over years of the intra-ensemble standard deviation (across all 20 members) at each lead. The dashed horizontal line marks a ratio of 1.0 (perfect calibration), with values below 1 indicating an under-dispersive (overconfident) ensemble and values above 1 indicating an over-dispersive ensemble. Orange shading in all panels marks the boreal spring months (March–May) associated with the spring predictability barrier (SPB).*

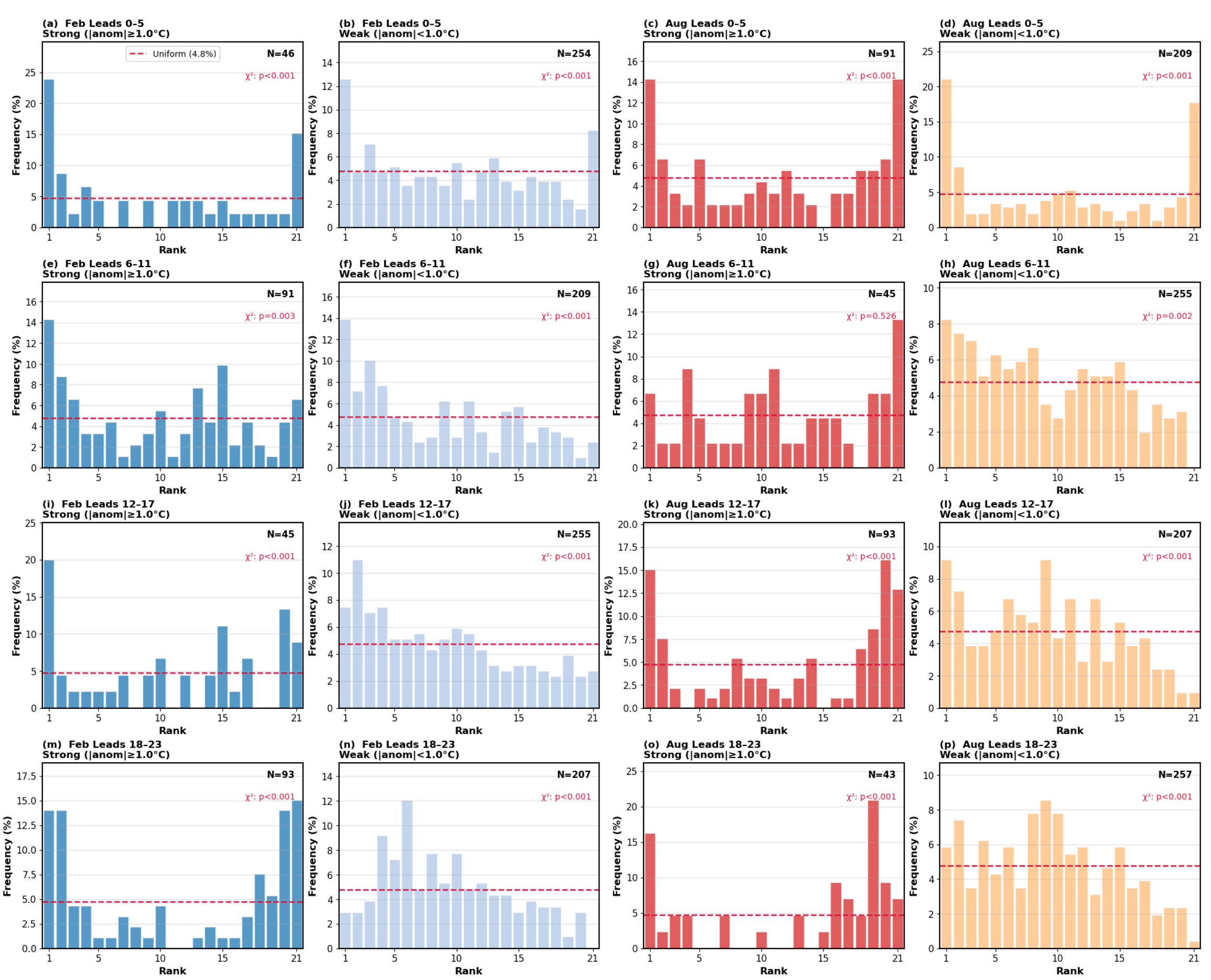

*Fig. 2. Conditional rank histograms for strong and weak ENSO events. Rank histograms are stratified by the magnitude of the observed Niño-3.4 anomaly at verification. Strong events (darker shading, left panel in each pair) are defined as cases where the verified observed anomaly satisfies $|\Delta T| \geq 1°C$; weak events (lighter shading, right panel) are all remaining cases. The four rows correspond to the aggregation of forecast leads of 0–5, 6–11, 12–17, and 18–23. The first two columns show the results of strong and weak ENSO events for forecasts initialized in February. The last two columns show the results for forecasts initialized in August. $\chi^2$ uniformity statistics are shown in each panel for reference only.*

## 4. Year-1 ENSO in February-Initialized Predictions

*a. Predictability and Uncertainty*

We next examine how ENSO predictability develops in February-initialized forecasts, using ESA to identify antecedent anomalies that constrain the December Niño-3.4. By interpreting the sensitivities alongside established Bjerknes feedback, recharge–discharge framework, and teleconnections (Bjerknes 1969; Jin 1997; Amaya 2019), we verify whether the ESA identifies physically plausible relationships.

The ESA reveals that SST and surface wind anomalies associated with December Niño-3.4 evolve in a manner consistent with known ENSO development mechanisms (Fig. 3). In boreal spring, the December Niño-3.4 index exhibits sensitivity to a subtropical SST pattern resembling the NPMM (Figs. 3a–b) (Vimont et al. 2001; Anderson 2003). This NPMM-like structure is characterized by off-equatorial SST and wind anomalies in the northeastern Pacific that influence ENSO through subtropical–tropical coupling (Amaya 2019; Wu et al. 2021). As the forecast progresses into April and May, the sensitivity strengthens and expands equatorward (Figs. 3c–d). Off-equatorial signals appear in both hemispheres. Alongside the NPMM-like structure, a southeastern subtropical pattern resembling the SPMM develops, consistent with its capacity to imprint on the equatorial Pacific through atmosphere-ocean coupling (H. Zhang et al. 2014). Both signals migrate toward the equator and merge by early summer, accompanied by growing sensitivity to westerly wind anomalies along the equator. This evolution is consistent with the meridional pathways by which the tropical precipitation and Hadley circulations couple subtropical and equatorial Pacific variability (Amaya et al. 2019; Li et al. 2023) (Supplementary Text). This temporal evolution marks a transition in forecast sensitivity from the NPMM-associated subtropical precursors to equatorial coupled dynamics as the Bjerknes feedback strengthens (Figs. 3e–j), consistent with previous studies (e.g., Amaya 2019).

The analysis above is restricted to the Pacific sector, but the same sensitivity fields computed over the global domain (Fig. S2) place the Pacific evolution in a pantropical context. At Lead 0 the sensitivities are weak and spatially diffuse across all basins, consistent with the limited ensemble spread near initialization (Section 2d). By Lead 4, the SLP signal in the tropical Indian Ocean becomes positive, and SST signals also emerge in the Indian Ocean and the tropical Atlantic. These signals suggest inter-basin coupling with the developing ENSO. By Lead 8, these sensitivities evolved and strengthened, manifesting a global pattern of ENSO teleconnections.

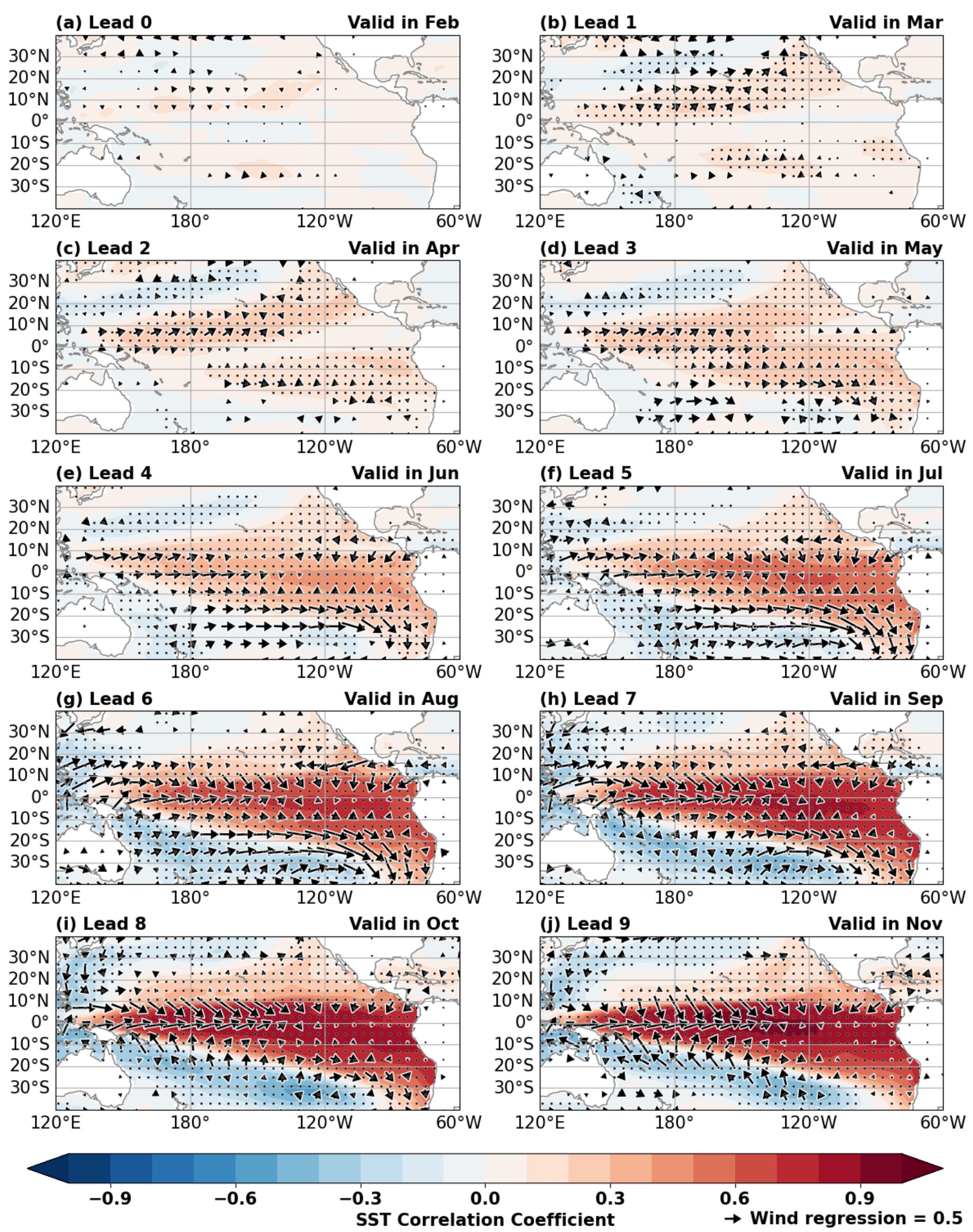

*Fig. 3. Ensemble sensitivity of the December Niño-3.4 index to SMYLE forecast states in February-initialized hindcasts. For each initialization year, the forecast SST anomaly at each grid point and lead is correlated across the 20 ensemble members with that year's December Niño-3.4 anomaly (Lead 10). The maps show these per-year correlation fields averaged over the 50 initialization years (1970–2019). Vectors show the corresponding member-wise regression of 10-m wind anomalies on the December Niño-3.4 index. Panels show lead months corresponding to (a) February, (b) March, (c) April, (d) May, (e) June, (f) July, (g) August, (h) September, (i) October, and (j) November. Stippling denotes grid points where the per-year*

*correlations are significantly sign-consistent across years at the 95% confidence level after Benjamini–Hochberg false discovery rate correction. Wind regression vectors are masked where neither the zonal nor meridional component is statistically significant at the 95% confidence level with two-tailed Student's t-test.*

The subsurface ESA shows that this transition is anchored in the equatorial thermocline (Fig. 4). We focus on temperature anomalies averaged over the 0–300 m layer, which serves a proxy for the variability of the equatorial Pacific thermocline (Zhao et al. 2021). Positive correlations in the eastern equatorial Pacific first become evident around May (Fig. 4c–d) and intensify through August (Figs. 4e–f), reflecting the tightening link between thermocline depth and the surface anomalies. The near-absence of subsurface correlations at initialization (Figs. 4a–b) is at least partly related to the atmosphere-only perturbations of SMYLE initialization (Section 2). The May emergence of ESA signals should be interpreted as an indicator of subsurface sensitivity becoming diagnosable. By November (Figs. 4g–h), a more pronounced zonal dipole emerges, characterized by strong positive correlations in the eastern Pacific and broad negative correlations to the west. These subsurface signals are the signature of coupled Bjerknes/recharge–discharge dynamics in which eastern-Pacific warming, weakened trade winds, and a zonally tilted thermocline reinforce one another (Bjerknes 1969; Jin 1997).

Taken together, the ESA patterns are consistent with established mechanisms of ENSO development, supporting ESA as a framework for identifying dynamically relevant precursor signals. The relatively weak and diffuse correlations at short leads may be attributed to the limited ensemble spread near initialization (Section 2d). Nonetheless, an alternative interpretation is that analyzing all years together obscures event-specific sensitivities. Accordingly, the next section turns to the years that SMYLE predicted best and worst.

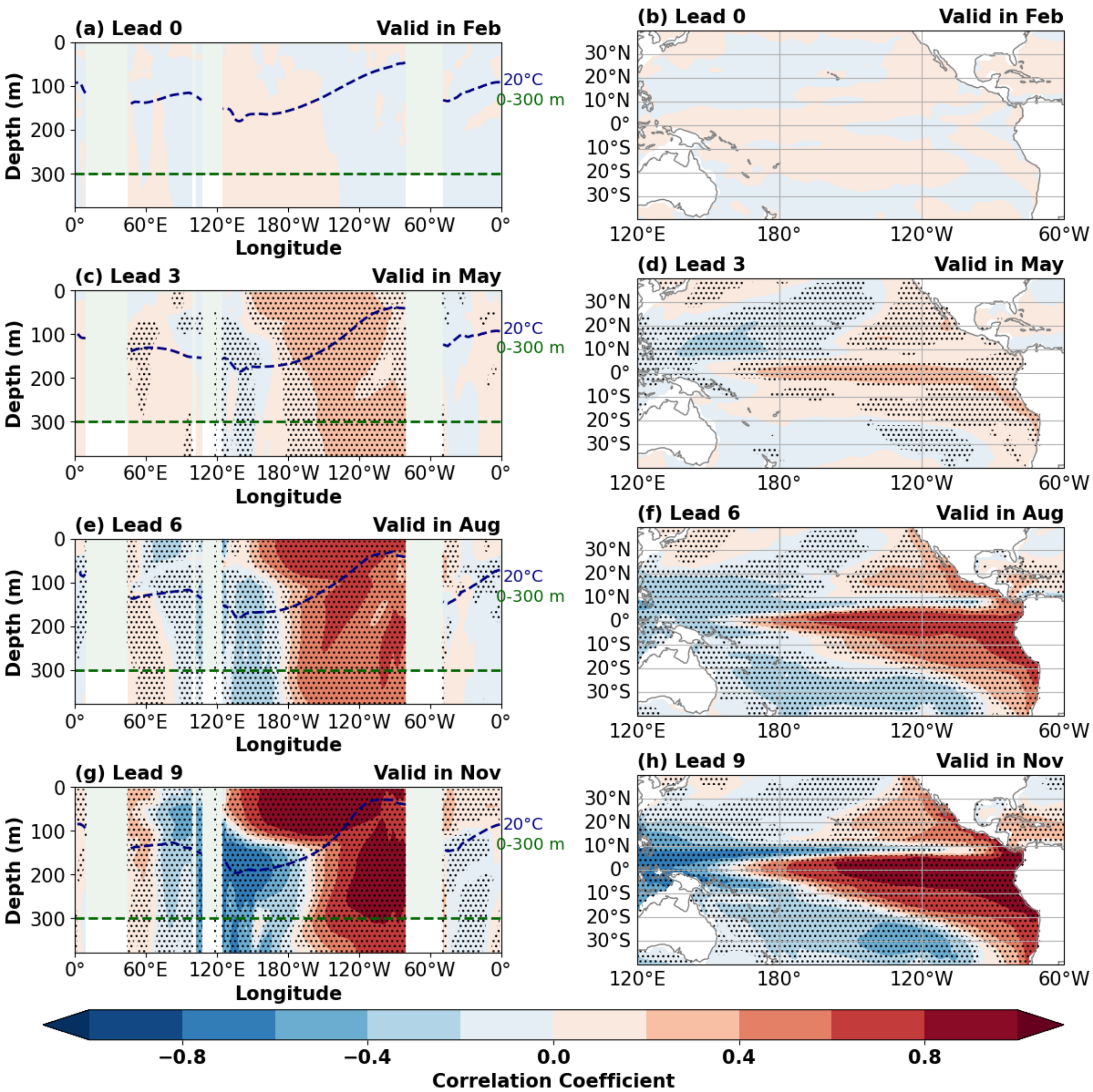

*Fig. 4. Ensemble sensitivity of the December Niño-3.4 index to SMYLE subsurface states in February-initialized hindcasts. For each initialization year, the forecast SST anomaly at each grid point and lead is correlated across the 20 ensemble members with that year's December Niño-3.4 anomaly (Lead 10). These per-year correlation fields are then averaged over the 50 initialization years (1970–2019). The left column shows equatorial (5°S–5°N) longitude–depth sensitivity, while the right column shows the sensitivity to subsurface temperature anomalies averaged over the 0–300 m layer. The dashed horizontal line in the left panels marks the 300 m reference level, and the navy dashed contour denotes the 20°C isotherm representing the thermocline. Panels correspond to different target months: (a, b) February, (c, d) May, (e, f) August, and (g, h) November. Shading indicates correlation coefficients, and stippling denotes correlations significant at the 95% confidence level.*

*b. Case Analyses of Best and Worst Predictions*

Because within-ensemble ocean variance degenerates at initialization and affects the efficacy of ESA (Section 2), initial state precursors are diagnosed from ensemble mean composites, which are independent of ensemble spread. Following the ranking defined in Section 2e, Fig. 5 examines how December Niño-3.4 forecast errors relate to the ENSO state. At initialization (Fig. 5a), errors show no significant overall dependence on the February state ($r = -0.11$), although among La Niña initializations the error does scale with the initial cold anomaly ($r = -0.62$, $p < 0.05$). At target time (Fig. 5b), errors become increasingly negative as the observed December anomaly increases ($r = -0.56$, $p < 0.001$), with a regression slope of −0.44. While the negative sign of the error-on-observation slope is expected for any predictive model that is not perfect, SMYLE forecasts are indeed disproportionately too weak when strong ENSO events occur (Fig. 2).

The extreme cases illustrate both the promise and the failure mode of the forecasts (Fig. 5 c-f). The SMYLE panels appear smoother than HadISST because ensemble averaging suppresses internal variability, with an additional contribution from the coarser 2° model grid. With those caveats in mind, we first examine the best prediction initialized in February 1982. Starting from a neutral initial state (−0.1 °C), SMYLE captured the developing 1982/83 extreme El Niño ten months ahead, with a December error of about −0.1 °C against an observed anomaly of +2.3 °C. The near-zero index error nonetheless conceals a structural pattern error. The prediction reproduced the equatorially centered warm-tongue structure of the observed event (Figs. 5c–d), but the predicted warm anomalies extend too far westward. This displacement is the anomaly-space signature of CESM2's excessive equatorial Pacific cold tongue, which steepens the climatological zonal SST gradient, strengthens the zonal advective feedback in the western Pacific, and shifts simulated ENSO SST anomalies westward relative to observations (Chen et al. 2021; Jiang et al. 2021). The worst prediction initialized in February 1994 shows a phase inconsistency: observations produced a moderate El Niño (Fig. 5f), whereas SMYLE predicted strong La Niña-like cooling in the Niño-3.4 region (Fig. 5e; error −2.6 °C).

Composite differences between the eight best- and worst-predicted years show that forecast success is imprinted on the initial state in a dynamically coherent way (Fig. 6). The best- and worst-predicted groups are comparable in the distribution of ENSO phase at the target: 1 El

Niño, 3 La Niña, and 4 neutral Decembers among the best-predicted years, versus 4, 2, and 2 among the worst. The difference is not statistically distinguishable (permutation $\chi^2$, p = 0.46), and the target amplitudes are likewise similar (mean |Niño-3.4| = 0.9 versus 1.1 °C, p = 0.69). At initialization, however, the two groups are well separated. The best-predicted years are drawn preferentially from La Niña February states (6 La Niña, 2 neutral, and 0 El Niño, versus 1, 6, and 1 among the worst; p = 0.04), with a mean February Niño-3.4 anomaly of −0.8 °C compared with +0.02 °C for the worst-predicted years (p = 0.02).

Consistent with the La Niña states, the best predicted years are significantly colder at the surface and in the eastern equatorial subsurface, yet significantly warmer in the western Pacific subsurface (Wsub box, Fig. 6b). The recharge oscillator theory (Jin 1997) suggests This recharged warm-pool configuration is associated with higher predictability. While the correlation between the prediction error and the initial ocean heat content is negative, the correlation across 50 initializations is insignificant, regardless of whether ORAS5 or SMYLE's own initial state are analyzed (not shown). This Pacific state is accompanied by significantly higher sea-level pressure over the northeastern Pacific and lower pressure over the Indian Ocean–Maritime Continent sector (Fig. 6c), which indicates enhanced easterly wind over the Pacific. As lead increases (Figs. 6d–i), robust differences develop in the NPMM and IOD regions while the Pacific signals persist, indicating that the initial La Niña state and the subtropical / inter-basin anomalies evolve together rather than independently.

The ESA composites computed separately for each group (Fig. S3) corroborate the findings from the sensitivity perspective. In the best-predicted years, sensitivity organizes along the equator by Lead 4 and forms a tight east–central Pacific structure by Lead 8 consistent with effective thermocline–SST coupling (Wyrtki 1975; Cane and Zebiak 1985; Jin 1997). In comparison, the worst-predicted years exhibit stronger but meridionally broad sensitivity extending deep into the subtropics of both hemispheres, indicating enhanced off-equatorial influences that interfere with the canonical equatorial development.

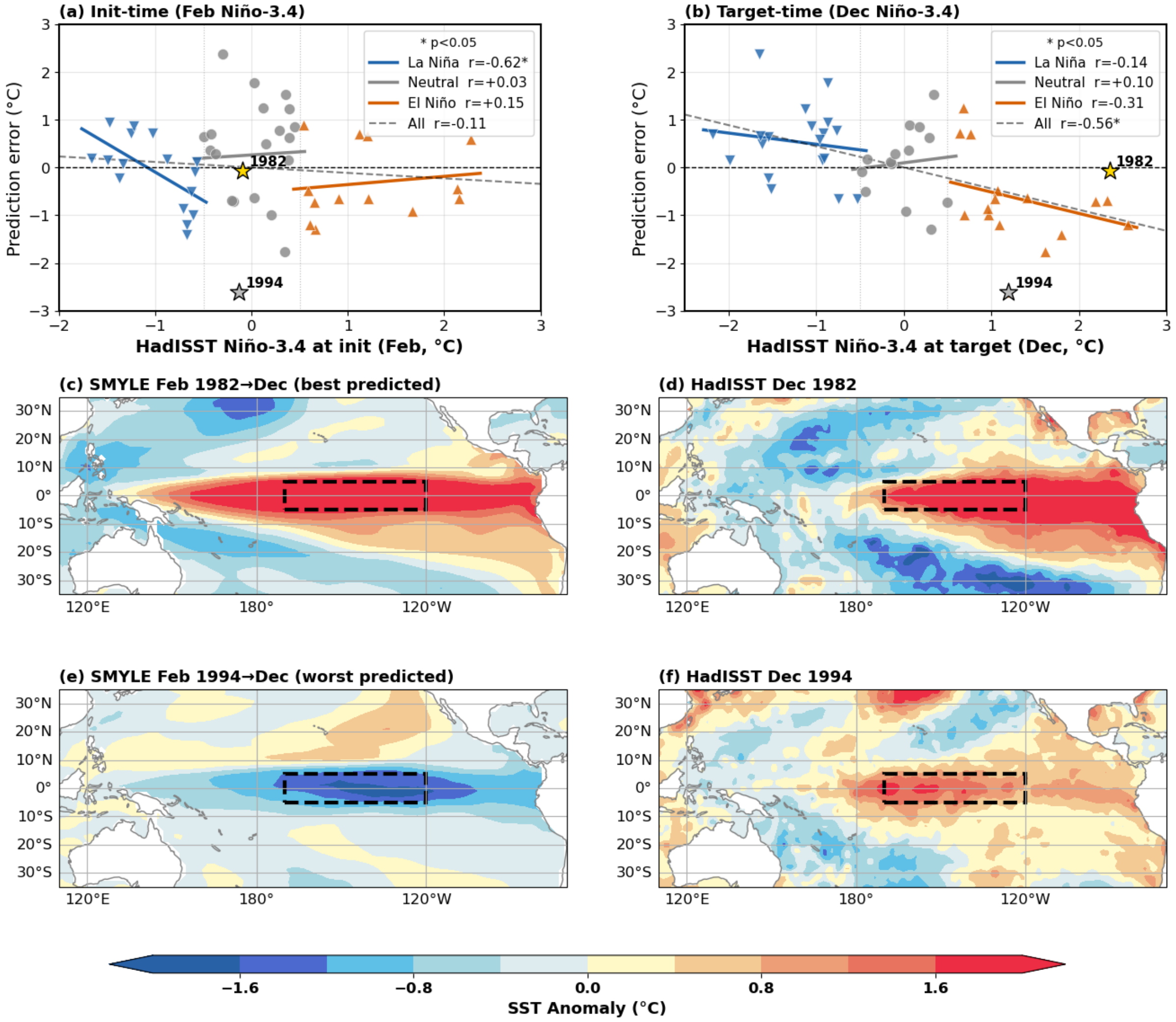

*Fig. 5. Errors of the SMYLE ensemble mean forecasts initialized in February in comparison with HadISST observations. (a) Scatterplot of December Niño-3.4 prediction error versus the HadISST Niño-3.4 anomaly at initialization time (February) for 1970–2019. (b) Same as (a), but the x-axis shows the HadISST Niño-3.4 anomaly at target time (December). In both panels, markers are colored by ENSO phase at the respective time: blue downward triangles for La Niña (Niño-3.4 < −0.5°C), gray circles for Neutral, and orange upward triangles for El Niño (Niño-3.4 > 0.5°C). Colored solid lines show within-phase regression fits; the dashed black line shows the overall regression. Correlation coefficients (r) are displayed in the legend; an asterisk indicates statistical significance at the 95% confidence level (p < 0.05). Gold and gray stars mark the best-predicted (1982) and worst-predicted (1994) years, respectively. (c, d) December SST anomalies for the best-predicted year (1982) from SMYLE (initialized in February) and HadISST, respectively. (e, f) Same as (c, d) but for the worst-predicted year (1994). The dashed black box marks the Niño-3.4 region (5°S–5°N, 170°–120°W).*

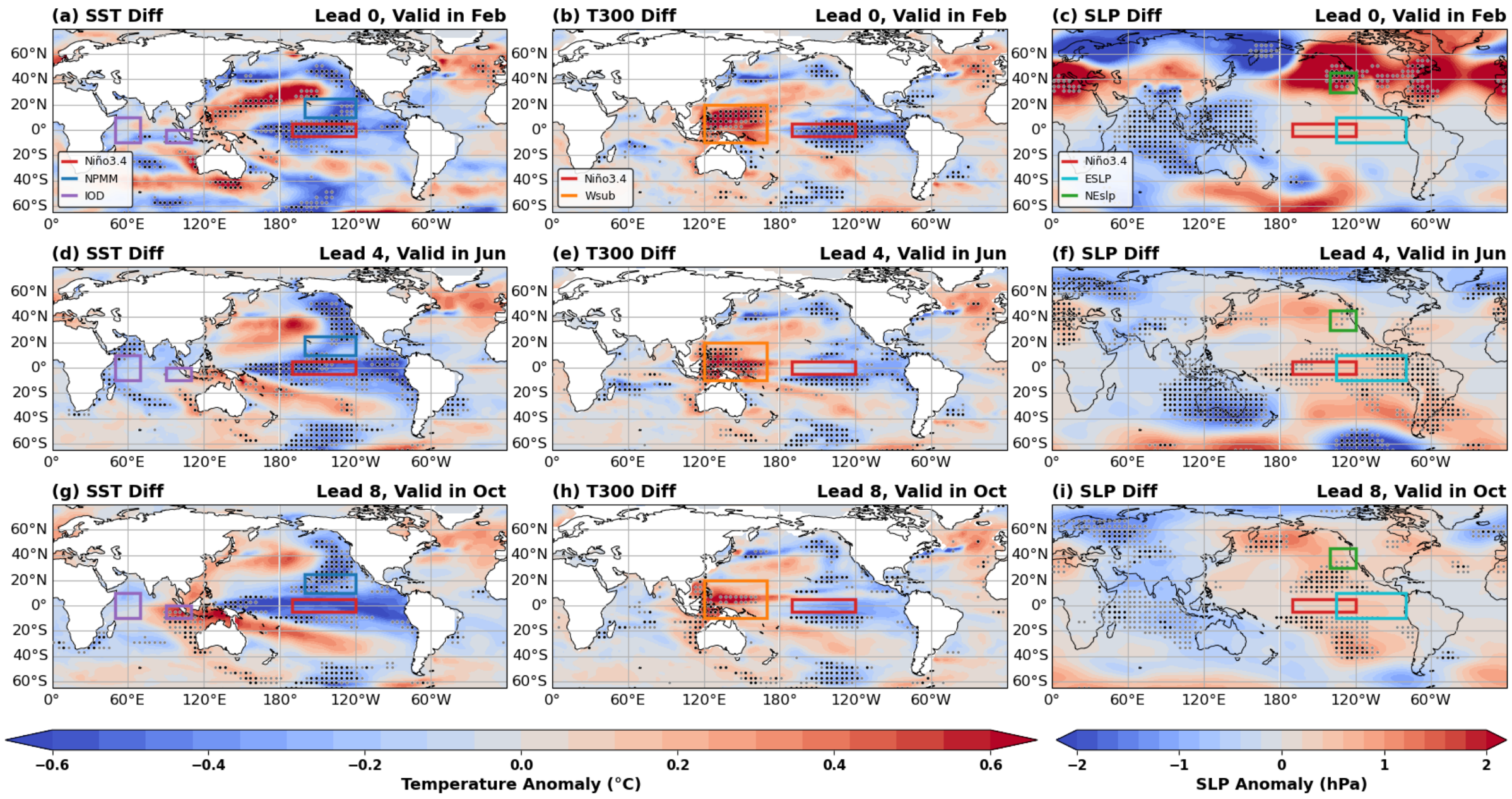

*Fig. 6. Difference maps of SST, subsurface temperature and SLP, defined as composites for the best-predicted years minus the worst-predicted years, based on February-initialized SMYLE forecasts for 1970–2019. Panels show lead times of (a, b, c) Month 0, (d, e, f) Month 4, and (g, h, i) Month 8. Red dashed rectangles indicate the Niño-3.4 region, colored boxes mark other key regions discussed in the text, and stippling denotes 95% significance.*

The regional indices (Fig. 7) quantify which initial state contrasts are robust. The best-predicted years show a cold surface state with a recharged Western Pacific subsurface and elevated Eastern Pacific SLP. This is the classic recharged La Niña state from which the recharge oscillator predicts a deterministic transition. This may explain why predictions initialized from the state are more skillful. Meanwhile, Northeastern Pacific SLP is significantly higher in the best years at initialization (~1.8 hPa; Fig. 7f), consistent with an anticyclonic anomaly on the southeastern flank of the Aleutian Low. This SLP pattern resembles a North Pacific Oscillation–type SLP anomaly, which forces subtropical SST anomalies through wind–evaporation–SST feedback and projects onto the NPMM (Ding et al. 2022; Liang et al. 2025). Meanwhile, Eastern Pacific SLP shows a weaker but positive contrast in early spring (Fig. 7e). Together they describe a coherent, elevated SLP pressure configuration in the East Pacific at the start of skillful forecasts. The predicted Niño-3.4 trajectories themselves separate significantly only in the first months (Fig. 7a), with best years initialized in colder ENSO states. The subtropical and cross-basin indices, by contrast, discriminate later in the forecast. The NPMM difference is persistently negative and becomes significant in late summer and autumn (Fig. 7b), and a negative IOD develops in the best-predicted years with significant separation from August to October (Fig. 7c), suggesting that these pathways reinforce, rather than initiate, the equatorial evolution set up by the initial recharge state.

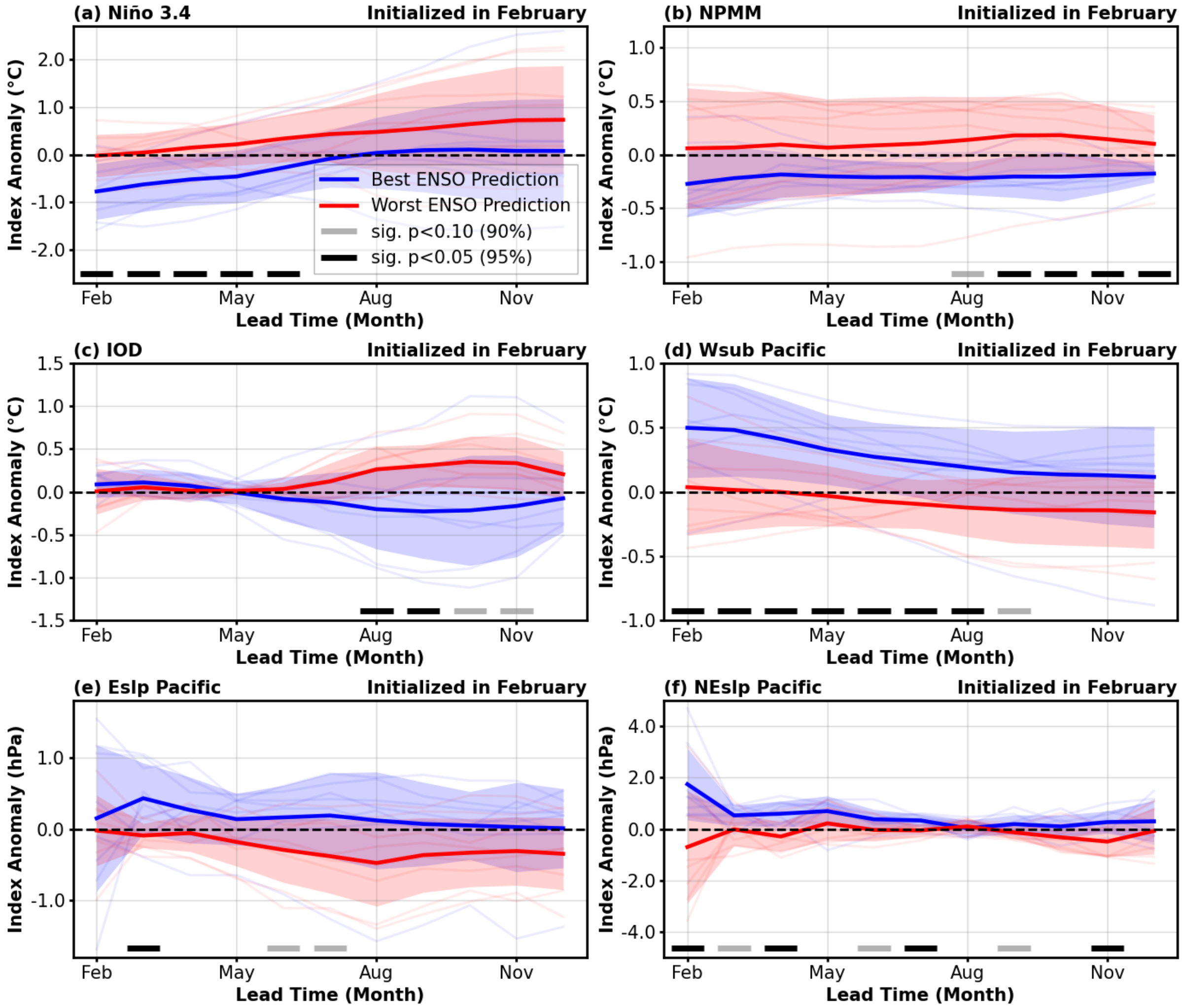

*Fig. 7. Panels (a)–(f) show anomaly time series of (a) the Niño-3.4 index and given regional indices (b) NPMM, (c) IOD, (d) Wsub, (e) ESLP, and (f) NEslp in the forecasts initialized in February. The x-axis indicates forecast lead in months. Blue lines denote the composite mean of the top eight best-predicted years, and red lines denote the composite mean of the top eight worst-predicted years, with light shading indicating ±1 standard deviation across composite years. Thin lines show individual-year ensemble mean time series for each composite group. Lead months with statistically significant differences between the best and worst groups based on a two-tailed Welch's t-test. indicated by horizontal bars at the bottom of each panel with gray bars for 90% confidence level and black bars for 95% confidence level.*

## 5. Year-2 ENSO in August-Initialized Predictions

### *a. Predictability and Uncertainty*

In contrast to the February-initialized predictions, in which subtropical precursors imprint on the equatorial Pacific within the first few months, predictability in August-initialized predictions emerges through a delayed pathway. Early lead-time signals are weak and spatially disorganized, and coherent structure develops only as ocean adjustment and large-scale circulation anomalies enable the Bjerknes feedback and other development mechanisms during Year-2 (Fig. 8). Subsurface ocean anomalies contribute to this adjustment process but do not emerge as the dominant predictor in the ESA results (Fig. 9). Since the linear ESA framework may understate nonlinear interactions, the interpretation of the lack of early Pacific precursors warrants caution.

During the first forecast autumn and winter, the ESA shows only scattered, weak correlations with the Year-2 December Niño-3.4 index (Figs. 8a–c). The subtropical NPMM footprint (Vimont et al. 2001; Amaya 2019) is faint and projects weakly onto the equator. A careful inspection of the global domain suggests weak wind signals in the Indian Ocean and the extratropical Pacific (Fig. S4). From about Lead 6 onward, NPMM-related signals emerge and are followed by eastward-amplifying equatorial SST correlations that become increasingly confined to the equatorial band through Lead 14. This is accompanied by strengthening westerly wind correlations over the central–western Pacific, the signature of an intensifying Bjerknes feedback (Figs. 8d–h).

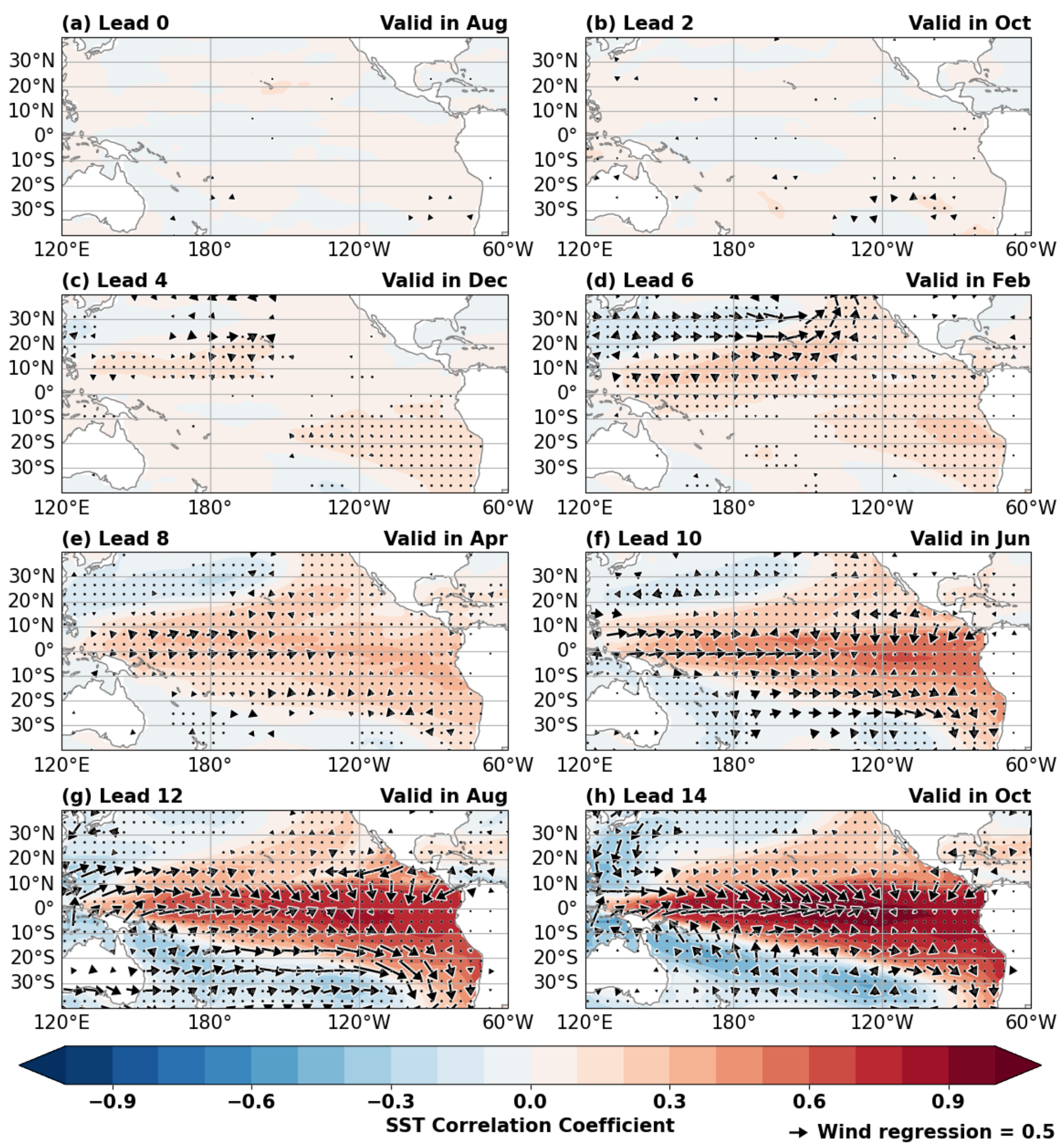

*Fig. 8. Ensemble sensitivity of the December Niño-3.4 index of the following year (Year+1) to SST anomalies, based on CESM2–SMYLE forecasts initialized in August over 1970–2019. Vectors denote the corresponding member-wise regression of surface wind anomalies (m s$^{-1}$ per °C) onto the Year+1 December Niño-3.4 index. Panels show anomaly fields valid in (a) August, (b) October, (c) December, (d) February of Year+1, (e) April, (f) June, (g) August, and (h) October of Year+1. Stippling indicates SSTA correlations that are statistically significant at the 95% confidence level. Wind regression vectors are masked where neither the zonal nor meridional component exceeds the 95% confidence level.*

The subsurface evolution is consistent with the emergence of Year-2 skill (Fig. 9). Correlations between subsurface temperature and the Year-2 December Niño-3.4 index remain weak and diffuse through the first forecast year (Figs. 9a–d). Part of this early weakness is again by SMYLE construction, given that August initializations the subsurface spread reaches half its saturated amplitude only by about Lead 7 (not shown). Therefore, the contrast between the early panels and the organized Year-2 dipole partly reflects the growth of diagnosable spread rather than purely the delayed organization of coupling. Statistically significant positive correlations first appear in the eastern equatorial upper thermocline (~100–200 m) by Lead 8 (April of Year-2; Figs. 9e–f). By Lead 12 (August of Year-2) they organize into a clear zonal dipole, with positive correlations in the east and negative correlations in the west (Figs. 9g–h), characteristic of recharge–discharge adjustment that subsequently projects onto surface ENSO anomalies (Wyrtki 1975; Jin 1997; Meinen and McPhaden 2000). The delayed consolidation of subsurface correlations mirrors the late emergence of coherent surface coupling in Fig. 8, contrasting with February forecasts in which precursor signals imprint on the equatorial subsurface within months.

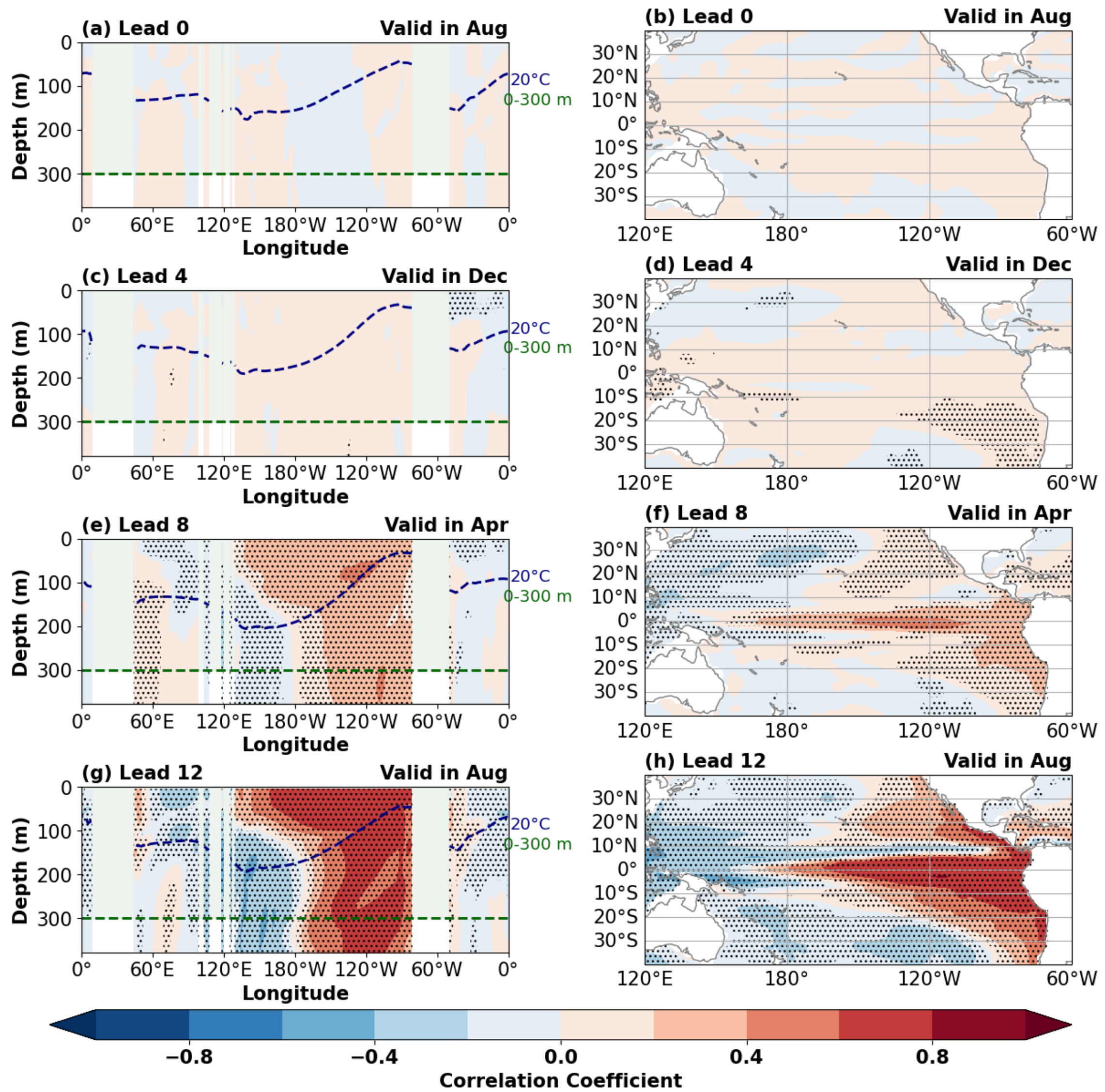

*Fig. 9. Ensemble sensitivity of the Year-2 December Niño-3.4 index to SMYLE subsurface temperature anomalies, based on forecasts initialized in August and averaged over 1970–2019. The left column shows equatorial (5°S–5°N) longitude–depth sensitivity, while the right column shows the sensitivity to subsurface temperature anomalies averaged over the 0–300 m layer. The horizontal dashed dark-green line in the left panels marks the 300 m reference level, and the dashed navy contour denotes the 20°C isotherm representing the thermocline. Panels correspond to different forecast months: (a, b) August, (c, d) December, (e, f) April of Year-2, and (g, h) August of Year-2. Stippling indicates correlations that are statistically significant at the 95% confidence level.*

### *b. Case Analyses of Best and Worst Predictions*

The amplitude-dependent bias identified for February-initialized predictions is stronger for the August-initialized predictions. Prediction errors show no significant correlations with the August initial state of Niño-3.4 (Fig. 10a; $r = -0.01$) but scale strongly with the observed Year-2 December anomaly ($r = -0.75$, $p < 0.05$) (Fig. 10b). The regression slope is −0.65 and substantially steeper than the February value of −0.44 (cf. Fig. 5b). When observed anomalies exceed about +2 °C, errors of forecasts cluster around −1 to −2 °C. While the ensemble averaging tends to weaken signals, the conditional rank histograms (Fig. 2) that the model under-predicts the amplitude of strong events. This damping of strong El Niño events may partly reflect underestimated coupled-feedback strength at long leads because of insufficient zonal wind stress responses to SST anomalies (Chen et al. 2021).

The best-predicted case demonstrates that skillful Year-2 forecasts are nonetheless achievable, even for strong events. Initialized in August 1983 from a nearly neutral state (Niño-3.4 ≈ −0.1 °C), SMYLE predicted the strong December 1984 La Niña sixteen months ahead with an error of only +0.04 °C, reproducing the equatorially confined cold-tongue structure seen in observations (Figs. 10c–d). The predicted equatorial anomalies, however, again extend too far westward, consistent with the systematic CESM2 pattern bias discussed in Section 4b. In comparison, the worst-predicted case (August 2009 → December 2010) missed the rapid transition from El Niño to strong La Niña. Initialized during a developing warm event (~0.7 °C), SMYLE maintained warm conditions into the second winter (forecast ~1.2 °C) while observations swung to about −1.6 °C, an error of ~2.8 °C with opposite-signed equatorial anomalies (Figs. 10e–f).

Unlike the February-initialized predictions, Year-2 forecast skill is selected by the amplitude of the outcome rather than by the initial state (Fig. 11). At initialization, the best- and worst-predicted groups are statistically indistinguishable: 2 La Niña, 5 neutral, and 1 El Niño among the best-predicted years, versus 1, 6, and 1 among the worst ($p = 1.00$). In comparison, the best-predicted Year-2 Decembers are overwhelmingly neutral (1 El Niño, 1 La Niña, and 6 neutral), whereas every worst-predicted year verifies in an active ENSO state (5 El Niño, 3 La Niña, and 0 neutral; $p = 0.01$). The August-initialized groups also differ in observed target amplitude (mean |Niño-3.4| = 0.4 versus 1.9 °C; $p < 0.001$). Overall, the amplitude-dependent bias at Year-2 leads (Fig. 10b) suggests extreme-error years are predominantly extreme-outcome or phase-transition years.

Composite analysis suggests the best-predicted years start with an El Niño-like thermocline configuration with a warmer eastern-Pacific upper ocean and cooler western-Pacific subsurface conditions (negative Wsub; Figs. 11a–b). The equatorial Pacific signals are accompanied by three significant remote signals: a positive IOD-like contrast in the Indian Ocean, significantly cooler SST in the tropical equatorial South Atlantic (TESA), and cooler SST in the extratropical Pacific region (Fig. 11a). These differences weaken, and in places reverse sign, as lead increases (Figs. 11d–l), indicating that the discriminating information resides in the initial and early forecast states. Interestingly, ESA of predictions initialized in August suggest best predictions are associated with Eurasia wind signals at Lead 6 (valid in Year-2 February), whereas worst predictions are related to pronounced NPMM-like signals (Fig. S5).

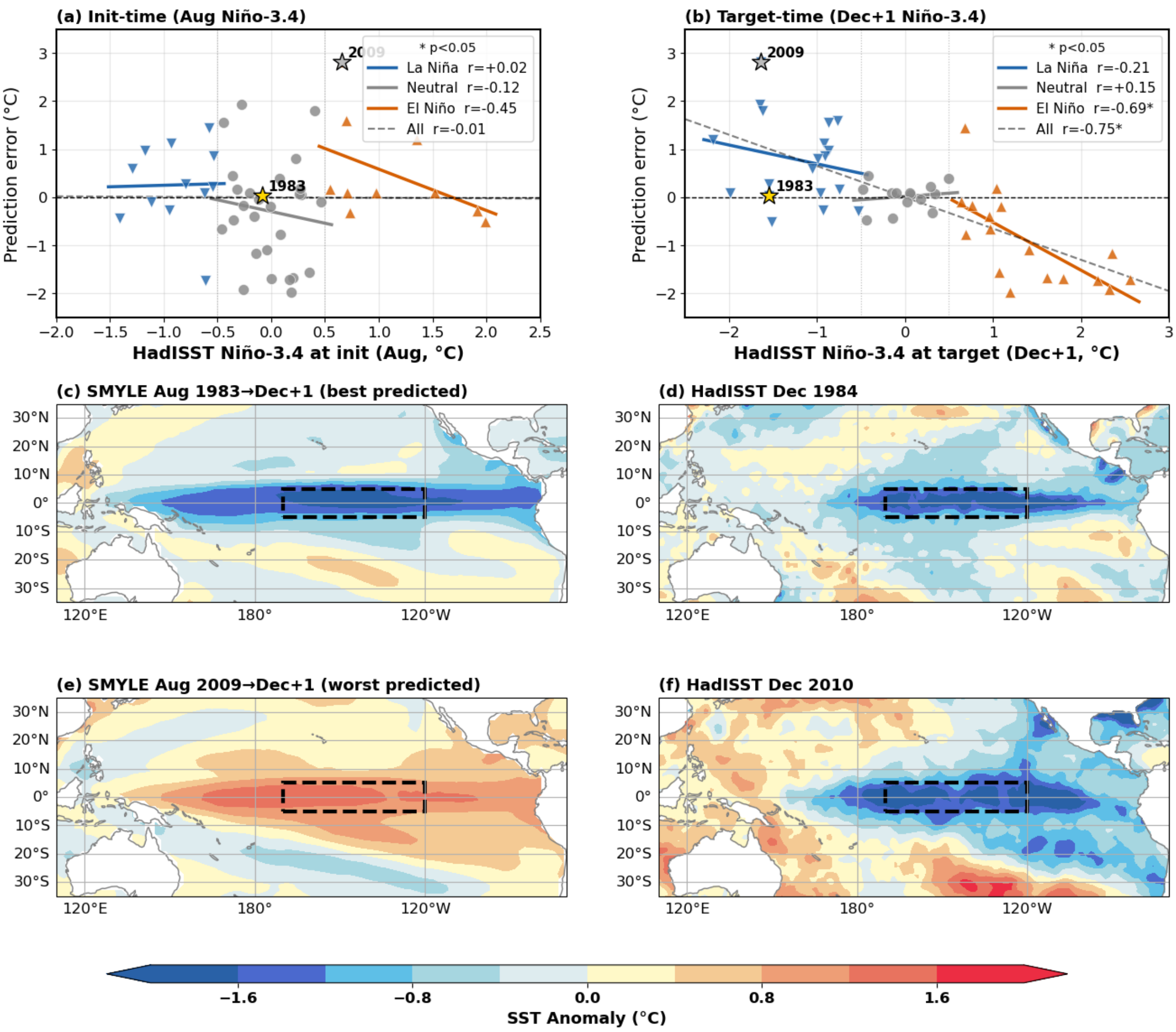

*Fig. 10. Errors of the SMYLE ensemble mean forecasts initialized in August in comparison with HadISST observations. (a) Scatterplot of the December (Year-2) Niño-3.4 prediction error — defined as the SMYLE ensemble mean forecast initialized in August minus the HadISST observation — versus the observed HadISST Niño-3.4 anomaly at initialization time (August), for each year during 1970–2019. (b) Same as (a) but versus the HadISST Niño-3.4 anomaly at the target time (December, Year-2). In both panels, scatter points are colored and shaped by*

*ENSO phase at the respective time (blue downward triangle: La Niña; gray circle: Neutral; orange upward triangle: El Niño), with within-phase regression lines and an overall black dashed regression line; correlation coefficients (r) for each phase and for all years are shown in the legend (asterisk indicates p < 0.05). A gold star marks the best-predicted year (1983) and a gray star marks the worst-predicted year (2009). (c, d) December SST anomalies for the best-predicted case (initialized August 1983, valid December 1984) from SMYLE and HadISST, respectively. (e, f) December SST anomalies for the worst-predicted case (initialized August 2009, valid December 2010) from SMYLE and HadISST, respectively.*

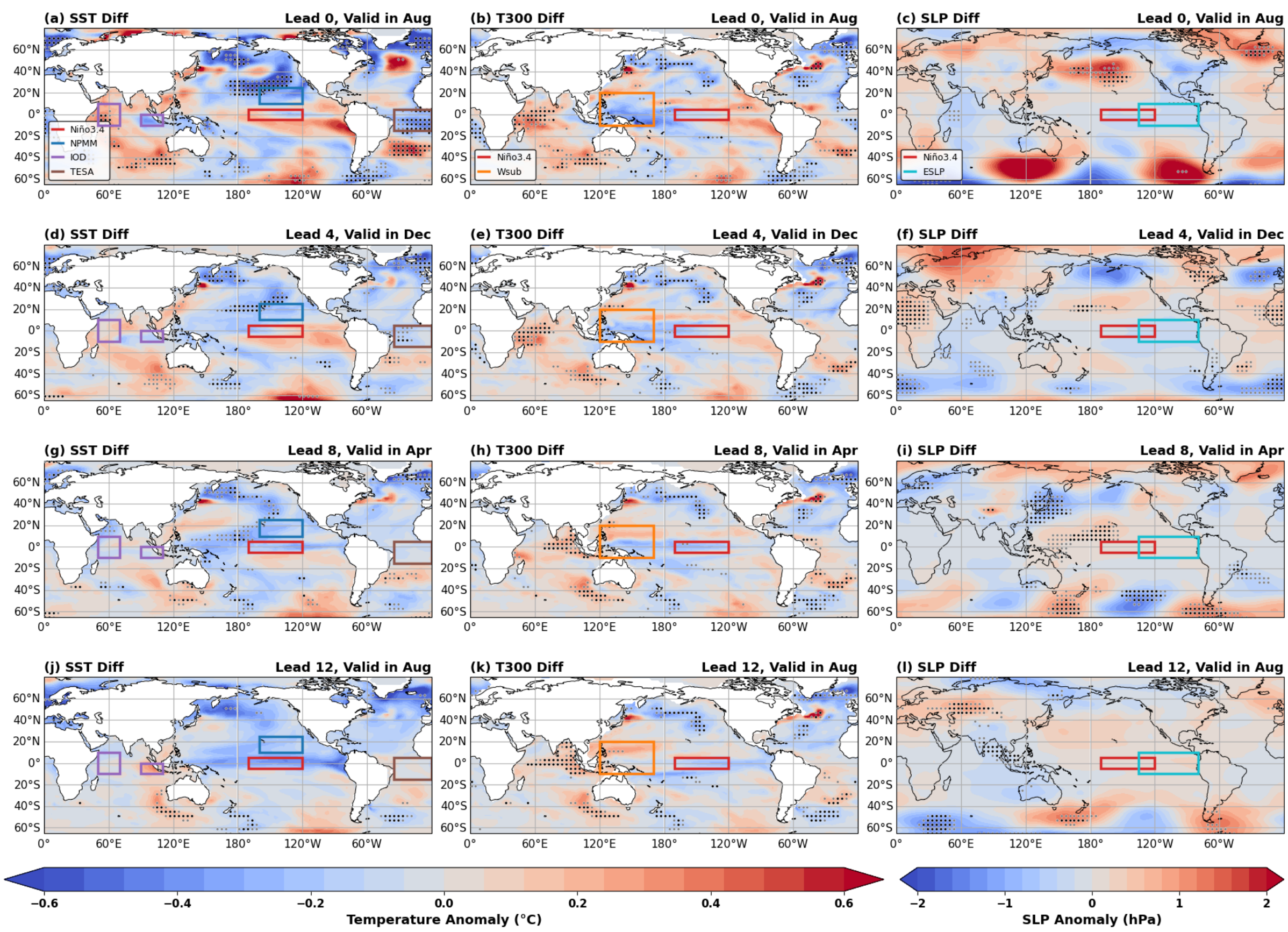

*Fig. 11. Difference maps of SST and subsurface temperature, defined as composites for the best-predicted years minus the worst-predicted years, based on August-initialized SMYLE forecasts for 1970–2019. Panels show lead times of (a, b, c) Month 0, (d, e, f) Month 4, (g, h, i) Month 8 and (j, k, l) Month 12. Red dashed rectangles indicate the Niño-3.4 region, colored boxes mark other key regions discussed in the text, and stippling denotes 95% significance.*

The regional indices further quantify which of these contrasts are robust (Fig. 12). The clearest early discriminator of best and worst predictions is the tropical equatorial South

Atlantic state (Fig. 12d), with the best-predicted years beginning with significantly cold anomalies ($p < 0.05$ at Leads 0–1). In observations, August TESA shows no significant correlation with the amplitude of the following Year-2 December outcome ($r = 0.08$, $p = 0.61$). The TESA contrast therefore does not arise from the outcome-amplitude imbalance between the groups. Independent support comes from the XRO ablation experiments, which attribute a small but resolvable skill contribution to TESA at the Lead 16 target ($\Delta ACC = +0.03$; Section 6). The two lines of evidence are consistent with the Atlantic influence on the ENSO (Rodríguez-Fonseca et al. 2009; Exarchou et al. 2021; L. Zhang et al. 2025; Jeffree et al. 2026). Positive IOD anomalies also characterize the best-predicted years in the first autumn (Fig. 12c; $p < 0.10$), consistent with the Indian Ocean's modulation of Pacific convection through atmospheric-bridge processes (Kug and Kang 2006; Izumo et al. 2010). The other indices show large differences but never reach significance. The Year-2 predictability of August forecasts thus appears to rest on inter-basin couplings (Indian and Atlantic Ocean), rather than on the Pacific subsurface and subtropical surface pathways that dominate Year-1 February forecasts.

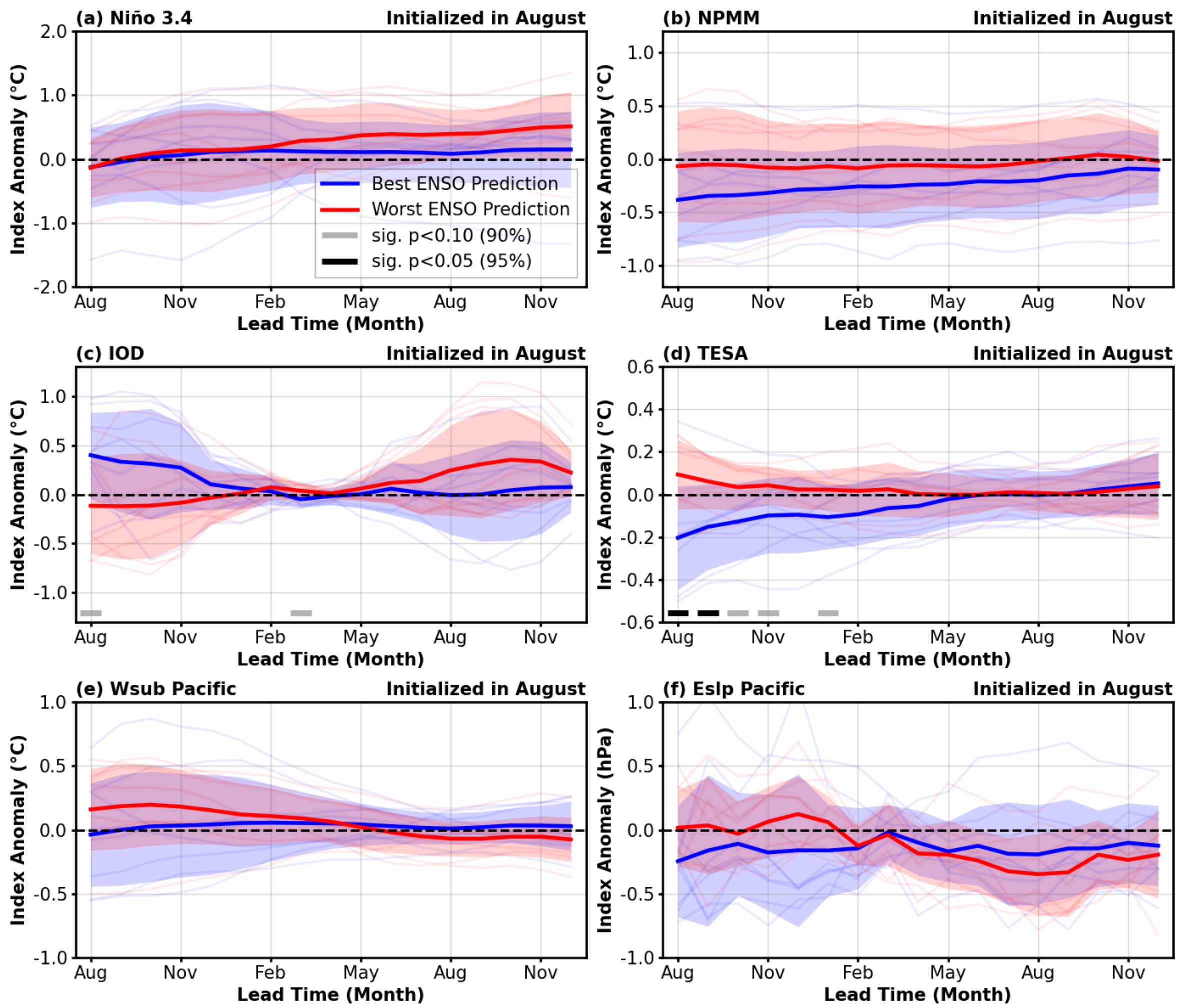

*Fig. 12. Panels (a)–(f) show anomaly time series of (a) the Niño-3.4 index and five regional indices defined in Table 1, all initialized in August. The x-axis indicates forecast lead time in months, labeled by the corresponding calendar month names. Blue lines denote the composite mean of the top eight best-predicted years, and red lines denote the composite mean of the top eight worst-predicted years, with light shading indicating ±1 standard deviation across composite years. Faint lines show individual year time series (ensemble means) for each composite group. Lead months with statistically significant differences between the best and worst groups based on a two-tailed Welch's t-test are indicated by horizontal bars at the bottom of each panel, with gray bars for 90% confidence level and black bars for 95% confidence level.*

## 6. Comparison of SMYLE and XRO

To assess whether SMYLE's coupled processes are consistent with the extended nonlinear recharge oscillator (XRO), we compare forecast skill (Fig. 13a–b), predictor-mode contributions (Fig. 13c–d), and process-level sensitivities (Fig. 14). For simplicity and consistency with previous studies (e.g., Zhao et al. 2024), the skill discussion here focuses on the ACC of ensemble mean December Niño-3.4 forecasts, computed identically for both systems.

When the ensemble size is matched at 20 members, the skill difference of SMYLE and XRO forecasts is narrow for the December targets (Fig. 13a–b). Specifically, the ACC difference between SMYLE and the full-fit XRO is up to ~0.02 at Lead 10 for February-initialized predictions and at Lead 16 for August-initialized predictions. A 200-member XRO raises the ACC values at those leads to ~0.70 and ~0.60, suggesting a sampling penalty of ~0.05 for 20-member forecasts. Importantly, the full-fit trains on data that overlap the forecast window: under leave-two-years-out cross-validation, the XRO's target-lead ACC drops by ~0.10 at both December targets. The XRO's cross-validated advantage mainly appears in February-initialized Year-2 leads and August-initialized Year-1 leads, where SMYLE's skill falls sharply through the spring predictability barrier. Consistent with Zhao et al. (2024), removing all remote predictor modes degrades the XRO at long leads (dotted curves in Fig. 13a–b), suggesting that including coupling with remote SST is valuable for long-lead skill.

The mode-ablation experiments identify which remote modes matter for the ENSO prediction at December targets (Fig. 13c–d). Collectively, the remote modes are essential.

Removing all of them drops ACC by ~0.20 in February-initialized predictions and ~0.16 in August-initialized predictions. This also degrades the XRO below SMYLE throughout Year-1. Individual attributions are subject to substantial sampling uncertainty at n = 50 years, but their ranking mirror the seasonal pathways of Sections 4 and 5. For February-initialized predictions, the NPMM is the leading contributor (ΔACC = +0.07), although its 95% confidence interval (CI) [−0.02, +0.17] spans across zero. Among the February modes, only TESA is formally distinguishable from zero with a small value (+0.01). For August-initialized predictions, the contributions shift to the inter-basin couplings. The IOB contribution is the largest (+0.07) but not separatable from zero (CI=[−0.01, +0.15]). In comparison, the contributions of IOD (+0.04, CI=[+0.00, +0.09]) and TESA (+0.03, CI=[+0.00, +0.05]) are separable from zero. Overall, the February result points to Pacific extratropical preconditioning, and the August result points to Indo-Atlantic coupling, consistent with the sensitivity structure diagnosed in SMYLE.

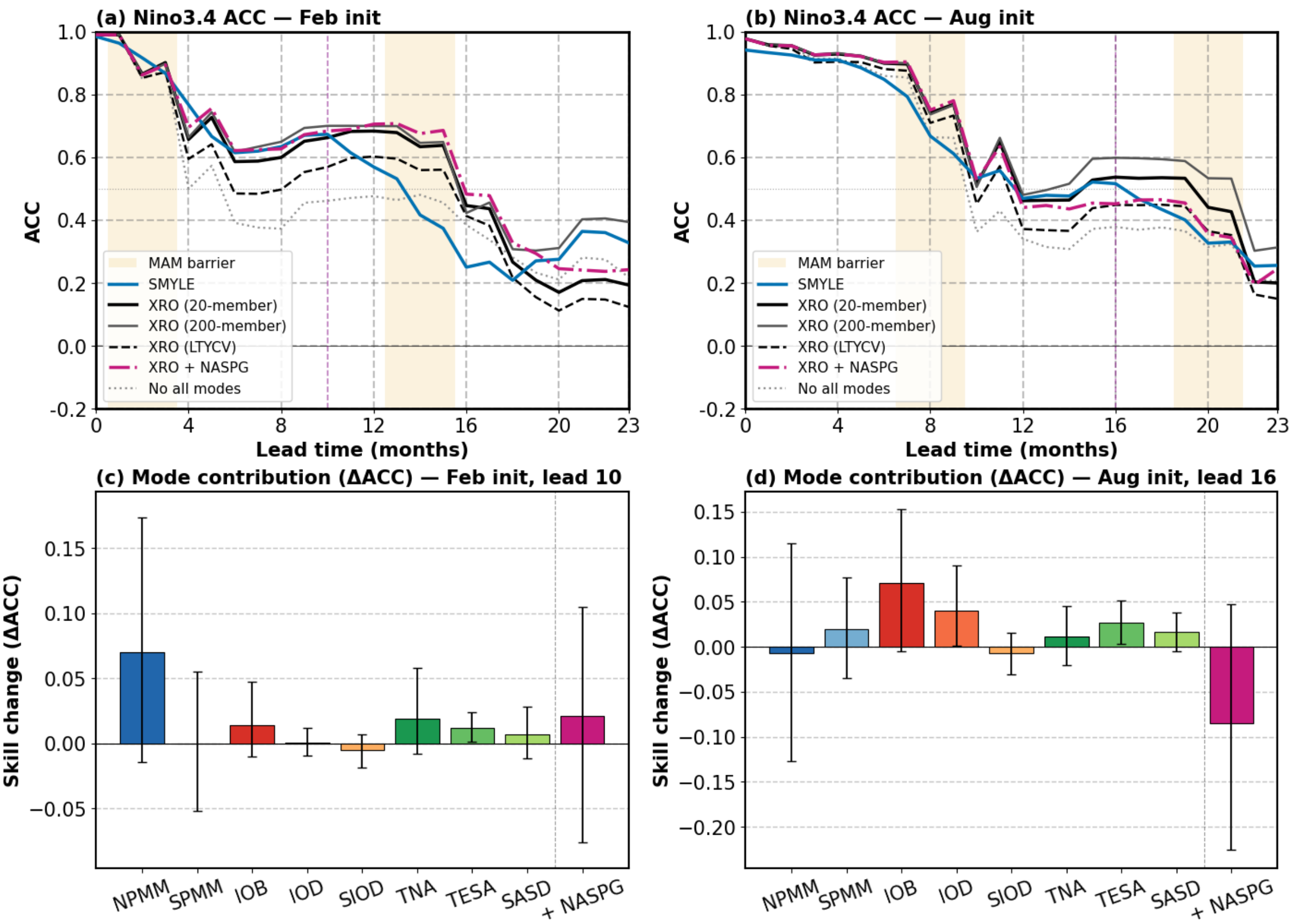

*Fig. 13. Niño-3.4 forecast skill and predictor-mode attribution for SMYLE and XRO hindcasts (1970–2019). Top row: anomaly correlation coefficient (ACC) of Niño-3.4 ensemble mean forecasts versus HadISST observations as a function of lead time, for (a) February and (b) August initialization. ACC is computed on linearly detrended anomalies. SMYLE (blue), XRO*

*with 20 ensemble members (black solid), XRO with 200 members (gray), XRO leave-two-years-out cross-validated skill (black dashed), and XRO augmented with the North Atlantic Subpolar Gyre (NASPG) index (pink dash-dot) are shown. Vertical yellow shading marks the boreal spring predictability barrier (MAM). Vertical pink lines indicate the target leads analyzed in (c)–(d): lead 10 (February → December Year +1) and lead 16 (August → December Year +1). Bottom row: skill change ($ACC_{full} - ACC_{ablated}$) at the target lead for (c) February and (d) August initialization. For the eight predictor ablations, positive ΔACC indicates that removing that mode degrades skill (i.e., the index leads to skill gain). The negative value range in (c) and (d) is set the same for visual consistency. Whiskers are 95% confidence intervals from a 5,000-draw bootstrap over verification years. The rightmost bar instead shows the skill change from adding the NASPG index (Supplementary Material).*

Applying the ESA to the same regional indices in SMYLE and XRO tests whether SMYLE develops predictor sensitivities with the observed timing (Fig. 14). The agreement is relatively strong for the Pacific modes. SMYLE reproduces the observed lead-dependence of NPMM sensitivity in the months preceding both December targets (Figs. 14a,i), even though the sensitivities are weaker than the XRO. The South Pacific Meridional Mode sensitivity also show a good agreement, with SMYLE indicating strong sensitivities before the December targets. The agreements for Indian Ocean indices are relatively good but seasonally dependent, with notable inconsistency around Lead 7 in February-initialized predictions and around Lead 13 in August-initialized predictions. More robust disagreements are present with the Atlantic couplings. SMYLE's February TESA sensitivity is nearly sign-opposed to XRO, and the negative TNA and TESA sensitivities that the XRO develops during Year-2 are much weaker in SMYLE (Figs. 14n,o).

The disagreement is noteworthy given that TESA is the index that most clearly discriminates SMYLE's best from worst August forecasts (Section 5b). CESM2 exhibits a substantial warm bias in the eastern equatorial Atlantic (Danabasoglu et al. 2020), reflecting a failure to maintain the cold tongue and its associated SST–wind–thermocline feedback. In SMYLE this bias is not merely inherited but actively re-emerges each forecast. The TESA-region mean state drifts warm by 1.0–1.4 °C within six months of initialization. We speculate that the challenge in representing the Atlantic state makes it difficult to simulate couplings with the ENSO realistically. This deficiency may degrade SMYLE skill when Atlantic forcings actively participate in the evolution of observed Year-2 ENSO events. While previous studies with other models suggest correcting SST variability in equatorial Atlantic improves the ENSO seasonal predication (Exarchou et al. 2021), a definitive conclusion for long-lead predictions requires pacemaker experiments that warrant future work.

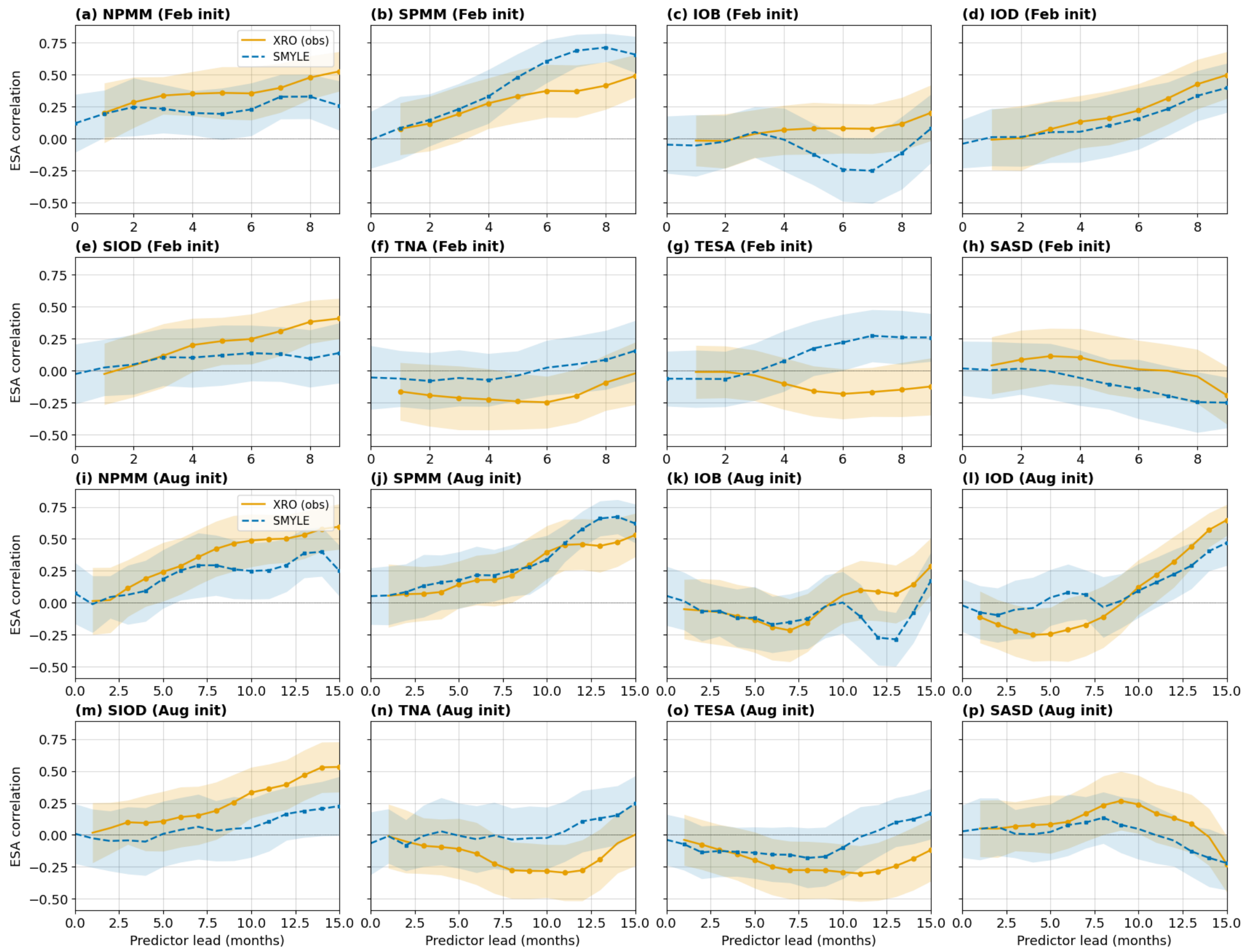

*Fig. 14. Ensemble sensitivity analysis (ESA) comparing the XRO observational benchmark and SMYLE CESM2-based hindcasts for Niño-3.4 predictability over 1970–2019. Each panel shows the ESA coefficient between a lagged regional-mean SST predictor index and the Niño-3.4 at the target verification month. Orange solid lines and shading denote the mean and ±1 standard deviation of the XRO forecasts years; blue dashed lines and shading denote the mean and ±1 standard deviation across SMYLE hindcast years. Filled circles (XRO) and squares (SMYLE) mark leads that are individually significant at $p < 0.05$ (two-tailed t-test). Panels (a)–(d) and (e)–(h) show the eight predictor modes for February initialization targeting December of Year +1 (lead 10); panels (i)–(l) and (m)–(p) show the same modes for August initialization targeting December of Year +1 (lead 16).*

## 7. Summary and Discussions

This study investigates precursors and pathways of ENSO predictability in the CESM2-SMYLE hindcast system. Using correlation-based ESA, we contrast February- and August-initialized predictions and benchmarking both against the extended nonlinear recharge oscillator (XRO) model. The key findings are as follows:

- SMYLE delivers meaningful ENSO skill at interannual leads, but with a state-dependent reliability problem. February-initialized predictions sustain ensemble mean ACC above 0.5 through Lead 13 (March of Year-2), and August-initialized predictions recover from a skill dip and retain ACC ≈ 0.5 for the December Year-2 target at Lead 16. Ensemble averaging is valuable to this performance, though marginal gains become small beyond about ten members. The probabilistic diagnostics, however, reveal that aggregate calibration conceals offsetting biases: the ensemble is severely overconfident for strong ENSO events while exhibiting a systematic warm bias for weak events. Nonetheless, the best-predicted cases in both initialization months demonstrate that some strong ENSO events can be predicted well by SMYLE.
- The pathways to predictability differ fundamentally between the two initialization months. For February-initialized predictions, ESA shows Year-1 predictability arising from Pacific processes. Signals associated with NPMM-like sensitivity in boreal spring migrate equatorward and join canonical Bjerknes feedback dynamics by early boreal summer. Skillful February forecasts are distinguished at initialization by a recharged western Pacific subsurface, organized Pacific SLP structures, and developing negative NPMM- and IOD-region anomalies. In contrast, low-skill forecasts exhibit meridionally broad patterns that indicate stronger sensitivities to extratropical anomalies. For August-initialized predictions, Year-2 predictability emerges through a delayed pathway with sensitivities remain weak and diffuse until equatorial coupled processes consolidate during Year-2. The initial state discriminators of best and worst predictions shift from the Pacific to inter-basin preconditioning, such as the significantly cold tropical equatorial South Atlantic (TESA) anomalies.
- SMYLE's coupled dynamics are largely consistent with the observationally constrained reduced-order benchmark despite notable disagreements regarding remote couplings. At matched ensemble size, SMYLE and the full-fit XRO are at near parity at the December targets. When evaluated against the cross-validated XRO, SMYLE shows higher skill at both targets. The XRO's advantage is nonetheless clear for February-initialized Year-2 leads and August-initialized Year-1 leads, where SMYLE's skill falls sharply through the spring predictability barrier. The XRO's mode attribution independently corroborates the two predictability pathways, with the NPMM leading contributions in February-initialized skill while the inter-basin couplings dominating August-initialized skill. At the process level, SMYLE reproduces the observed sensitivities for most Pacific and Indian Ocean modes but struggles with the Atlantic couplings.

Several caveats should guide interpretation. First, averaging ESA across years can obscure event-specific sensitivities associated with ENSO diversity, as the best/worst stratification demonstrates. Second, some biases may partly reflect initialization shock and lead-dependent drift toward the preferred climatology of CESM2. Whether such state-dependent behaviors are model-specific or characterize realistic ENSO dynamics warrant further investigation. Third, as a linear diagnostic, ESA cannot fully represent nonlinear or state-dependent pathways in real-world ENSO dynamics or simulations by SMYLE and XRO. The 50-year hindcast record limits a robust skill evaluation for the most extreme ENSO events. While conducting sensitivity experiments with SMYLE is computationally expensive, a fully nonlinear treatment such as adjoint sensitivities of the fitted XRO operator is a natural extension.

Lastly, the comparison of SMYLE and XRO seeks to improve physical understanding rather than to encourage optimization for arbitrary benchmarks. A fair comparison is inherently challenging. The XRO is initialized from the same ORAS5 analysis to which it is fitted, so its initial conditions are essentially perfect within its own state space. Conversely, the XRO is trained toward ORAS5 indices but verified against HadISST, which result in a small dataset-consistency penalty. In comparison, SMYLE carries initialization shock and lead-dependent drift that understates the full potential of the dynamical system. Rather than focusing skill metrics, it is more valuable to explore how the two models may complement each other, as demonstrated by the exploration of NASPG and TESA indices in Section 6.

Our findings carry practical implications for extending ENSO prediction to interannual leads. For February-initialized first-year skill, the priorities are accurate initialization of the Pacific state, such as the western Pacific subsurface and the subtropical NPMM state. This finding from the SMYLE hindcast is consistent with that from the NMME hindcasts (Zhang 2023). For August-initialized second-year skill, the key for skill improvements lies in the inter-basin couplings, especially improving CESM2's representation of Indian Ocean Basin and tropical Atlantic interactions with the Pacific. Lastly, ESA suggests additional extratropical precursors (e.g., NASPG) can affect the ENSO prediction. Future research should explore these relationships as well as other applications of ESA.

*Acknowledgments*

Y.M. and G.Z. thank Profs. Ryan Sriver and Zhuo Wang for stimulating discussions on the initial results. G.Z. acknowledges the support by the U.S. National Science Foundation (NSF) award RISE-2530555, as well as the faculty development fund of the University of Illinois Urbana-Champaign. S.Z. thanks the support by the U.S. NSF award 2334306.

*Data Availability Statement*

The HadISST monthly sea surface temperature (SST) data were obtained from the Met Office Hadley Centre and are openly available at https://www.metoffice.gov.uk/hadobs/hadisst/. The CESM2 Seasonal-to-Multiyear Large Ensemble (SMYLE) prediction experiment outputs were accessed directly from NCAR's high-performance computing environment (DOI: 10.26024/PWMA-RE41). XRO code is available at Github (https://github.com/senclimate/XRO). The analysis code will be made public via Zenodo repository before publication.